\documentclass[a4paper,11pt]{article}

\usepackage{jheppub} 
\usepackage{graphicx,color}
\usepackage{amsmath}
\usepackage{slashed}
\usepackage{multirow}
\usepackage{autobreak}
\usepackage[normalem]{ulem}
\usepackage{hyperref}
\usepackage{cleveref}
\usepackage{slashed}
\usepackage{amssymb}
\usepackage{epsfig}
\usepackage{array}
\usepackage{mathtools}
\usepackage{caption}
\usepackage[utf8]{inputenc}
\usepackage{mathtools, nccmath}
\usepackage{float}
\usepackage[section]{placeins}
\usepackage[utf8]{inputenc}
    \usepackage[T1]{fontenc}
    \usepackage{lmodern}
    \usepackage[x11names]{xcolor}
    \usepackage{framed}
\colorlet{shadecolor}{blue!10}

\SetSymbolFont{largesymbols}{bold}{OMX}{txex}{b}{n}

\DeclareMathAlphabet{\mathpzc}{OT1}{pzc}{m}{it}

\newcommand{\be}{\begin{eqnarray*}}
\newcommand{\ee}{\end{eqnarray*}}

\usepackage{parskip}

\newcommand{\ba}{\begin{array}}
\newcommand{\ea}{\end{array}}
\newcommand{\bd}{\begin{displaymath}}
\newcommand{\ed}{\end{displaymath}}
\newcommand{\besub}{\begin{subequations}}
\newcommand{\eesub}{\end{subequations}}

\def\q2 {q^2}

\newcommand{\lsim}{\raisebox{-0.13cm}{~\shortstack{$<$ \\[-0.07cm]
      $\sim$}}~}

\def\bt{\begin{table}}
\def\et{\end{table}}

\usepackage{bbm}  
\usepackage{amsmath}                                                

\newcommand{\nc}{\newcommand}
\nc{\beq}{\begin{equation}}  \nc{\eeq}{\end{equation}}
\nc{\bea}{\begin{eqnarray}}  \nc{\eea}{\end{eqnarray}}
\nc{\baa}{\begin{array}}     \nc{\eaa}{\end{array}}
\nc{\bit}{\begin{itemize}}   \nc{\eit}{\end{itemize}}
\nc{\ben}{\begin{enumerate}} \nc{\een}{\end{enumerate}}
\nc{\bce}{\begin{center}}    \nc{\ece}{\end{center}}
\nc{\bpm}{\begin{pmatrix}}   \nc{\epm}{\end{pmatrix}}
\nc{\bvt}{\begin{verbatim}}  \nc{\evt}{\end{verbatim}}
\nc{\bal}{\begin{align}}
\def\to{\rightarrow}

\def\boldoverdot{\,{\raise6pt\hbox{\bf.}\!\!\!\!\>}}

\def\ee{{\bf e}}

\usepackage{bm}
\def\diag{\hbox{\diag}}

\def\doubleundertext#1{
{\undertext{\vphantom{y}#1}}\par\nobreak\vskip-\the\baselineskip\vskip4pt%
\undertext{\hbox to 2in{}}}
\def\inbox#1{\vbox{\hrule\hbox{\vrule\kern5pt
     \vbox{\kern5pt#1\kern5pt}\kern5pt\vrule}\hrule}}
\def\sqr#1#2{{\vcenter{\hrule height.#2pt
      \hbox{\vrule width.#2pt height#1pt \kern#1pt
         \vrule width.#2pt}
      \hrule height.#2pt}}}

\def\today{\ifcase\month\or
  January\or February\or March\or April\or May\or June\or
  July\or August\or September\or October\or November\or December\fi
  \space\number\day, \number\year}
\def\pmb#1{\setbox0=\hbox{#1}%
  \kern-.025em\copy0\kern-\wd0
  \kern.05em\copy0\kern-\wd0
  \kern-.025em\raise.0433em\box0 }

\def\pmbb#1{\setbox0=\hbox{#1}%
  \kern-.02em\copy0\kern-\wd0
  \kern.04em\copy0\kern-\wd0
  \kern-.02em\raise.03464em\box0 }
\def\sumprime_#1{\setbox0=\hbox{$\scriptstyle{#1}$}
  \setbox2=\hbox{$\displaystyle{\sum}$}
  \setbox4=\hbox{${}'\mathsurround=0pt$}
  \dimen0=.5\wd0 \advance\dimen0 by-.5\wd2
  \ifdim\dimen0>0pt
  \ifdim\dimen0>\wd4 \kern\wd4 \else\kern\dimen0\fi\fi
\mathop{{\sum}'}_{\kern-\wd4 #1}}
	\title{\boldmath Probing anomalous $Z\,Z\,\gamma$ and $Z\,\gamma\,\gamma$ Couplings at the $e^+\,e^-$ Colliders using Optimal Observable Technique}

\author[a]{Sahabub Jahedi,}
\author[b]{Jayita Lahiri}

\affiliation[a]{Department of Physics, Indian Institute of Technology, Guwahati, Assam 781039, India}
\affiliation[b]{Institut f{\"u}r Theoretische Physik, Universit{\"a}t Hamburg, Luruper Chaussee 149, 22761 Hamburg, Germany}

\emailAdd{sahabub@iitg.ac.in}
\emailAdd{jayita.lahiri@desy.de}

\abstract{We study the anomalous $ZZ\gamma$ and $Z\gamma\gamma$ couplings that can be probed via $Z\gamma$ production at the $e^+ \, e^-$ colliders. We take Standard Model Effective Field Theory (SMEFT) approach to examine these anomalous neutral triple gauge couplings in a model independent way. There are four independent dimension-8 operators that generate these gauge interactions, one of them is CP-conserving and rest three are CP-violating. We adopt optimal observable technique  to extract the sensitivity at which these anomalous couplings can be probed at future $e^+e^-$ colliders and then compare the results with the latest experimental limit obtained at the LHC. We also study the impact and advantage of beam polarization in these precision measurements. Statistical limit (95\% C.L.) on individual anomalous couplings as well as the correlation between various couplings have been discussed in detail.}


\DeclareUnicodeCharacter{2212}{-}

\begin{document}

\maketitle
\flushbottom

\section{Introduction}
\label{sec:intro}

The Standard Model (SM) is currently being tested to a large degree of precision in various experiments, especially in the Higgs and flavor sector. The gauge sector of SM is equally predictive as well as restrictive because of the gauge symmetry $SU(2)_L\times U(1)_Y$ of SM. The interactions between the charged gauge boson ($W^{\pm}$) with the neutral gauge bosons ($\gamma, Z$) have been measured with great precision~\cite{L3:2004ulv,CMS:2013ant,CMS:2019ppl}. However, the self-interactions among neutral gauge bosons are absent within SM. Therefore, measurement of the self-interactions between the gauge bosons can be a probe of the gauge structure of SM. A deviation from the SM prediction in this sector will be a possible hint towards New Physics (NP) beyond SM. 

In this context, the production of $e^+e^- \rightarrow Z\gamma$ process has received significant attention, since the past few decades. This process takes place via $t-$channel electron exchange in the SM. The $s$-channel contribution via $Z$/$\gamma$ mediation is forbidden in the SM at tree-level, since neutral triple gauge-couplings $Z\gamma\gamma$ or $ZZ\gamma$ are not allowed at the tree level under the gauge symmetry $SU(2)_L \times U(1)_Y$. The loop-induced $s$-channel contribution within SM is also largely suppressed. Therefore, any deviation in the measurements in the process $e^+e^-\rightarrow Z\gamma$ from their SM prediction, will be useful hint for NP.  
$Z \gamma$ production  has been extensively studied in the literature in the context of the $e^+ \, e^-$ collider \cite{Choudhury:1994nt,Atag:2004cn,Ots:2004hk,Ots:2006dv,Gutierrez-Rodriguez:2008rvy,Ananthanarayan:2011fr,Ananthanarayan:2014sea,Rahaman:2016pqj,Rahaman:2017qql,Ellis:2019zex,Ellis:2020ljj} as well as $pp$ collider \cite{Baur:1992cd,Ellison:1998uy,Senol:2018cks,Yilmaz:2019cue,Senol:2019qyl,Yilmaz:2021ule,Rahaman:2018ujg,Ellis:2022zdw}.

In this work, we will follow the approach taken by various earlier works~\cite{Hagiwara:1986vm, Renard:1981es, Gounaris:1983zn, Baur:1992cd, Gounaris:1996rz, Gounaris:1999kf}, expressing anomalous neutral triple gauge-couplings (aNTGCs) in terms of model-independent dimension-6 and -8 operators~\footnote{Studies of aNTGCs regarding flavor-changing-neutral-current(FCNC) have been performed in~\cite{Hernandez-Juarez:2021mhi,Hernandez-Juarez:2022kjx}.}. The dimension-8 operators give rise to aNTGCs at tree-level where dimension-6 operators can lead to aNTGCs at one-loop level. We would like to state that in the present work, we focus only on the dimension-8 operators. However, it should be kept in mind that with sufficiently high NP scale, the effect of loop-induced aNTGC vertices arising from dimension-6 operators start making the dominant contribution. Following~\cite{Hagiwara:1986vm} in particular, we write the Lagrangian for aNTGCs in terms of dimension-8 couplings $h_i^V$, where $i=1,..4$ and $V=\gamma,Z$. $i=1,2$ denote CP-violating dimension-6 and -8 couplings, whereas $i=3,4$ denote CP-conserving dimension -6 and -8 couplings respectively.     

We must understand that, the objective of precision measurement of aNTGCs naturally drives us towards a lepton collider machine, on the one hand via the elimination of the QCD background and PDF uncertainties of hadron collider, on the other hand through the advantage of initial beam polarization to enhance the NP signal and/or suppress SM background. 
Therefore, we will focus our analysis on the proposed linear colliders such as International Linear Collider (ILC) \cite{Behnke:2013lya} and Compact Linear Collider (CLIC) \cite{Aicheler:2018arh}. However, our results must be pitted against the existing best sensitivity obtained at the LHC~\cite{ATLAS:2011nmx,ATLAS:2013way,CMS:2015wtk,CMS:2016cbq,ATLAS:2018nci} as well as LEP~\cite{L3:2004hlr,OPAL:2003gfi} which is another $e^+e^-$ collider, and therefore a precision machine. Comparing all the aforementioned results, we find that the most stringent bound so far comes from ATLAS~\cite{ATLAS:2018nci}, at $\sqrt{s}=13$ TeV and 36.1 $\rm fb^{-1}$ integrated luminosity. Therefore, we will compare this set of bounds with our prediction for upcoming ILC/CLIC.   

In this work, we will consider $e^+\,e^- \rightarrow Z\gamma$ at ILC with $\sqrt{s}=1$ TeV and CLIC with $\sqrt{s}=3$ TeV and an integrated luminosity ${\cal L}=$1000 $\rm fb^{-1}$, and determine the optimal statistical precision that can be achieved in those experimental analyses, in measuring the anomalous triple gauge couplings, using  optimal-observable technique (OOT)~\cite{Atwood:1991ka, Davier:1992nw, Diehl:1993br, Gunion:1996vv}. We will also explore the effect of beam polarization in the precision measurement in the context of both ILC and CLIC. In the past, OOT has been widely used in precise estimation of Higgs couplings~\cite{Hagiwara:2000tk,Dutta:2008bh} and top-quark couplings~\cite{Grzadkowski:1996pc,Grzadkowski:1997cj,Grzadkowski:1998bh,Grzadkowski:1999kx,Grzadkowski:2000nx} in the context of ILC, probing top-quark interactions in $\gamma\gamma$ collider~\cite{Grzadkowski:2003tf,Grzadkowski:2004iw,Grzadkowski:2005ye}, determination of CP properties of Higgs boson in top-Yukawa couplings at LHC~\cite{Gunion:1998hm} as well as at muon collider~\cite{Hioki:2007jc} and $e\gamma$ collider~\cite{Cao:2006pu}. This technique has also been used in probing heavy charged fermions at $e^+e^-$ collider~\cite{Bhattacharya:2021ltd} and also in studying NP effect in the context of flavor physics  \cite{Bhattacharya:2015ida,Calcuttawala:2017usw,Calcuttawala:2018wgo}.

This paper is organized as follows: In Section~\ref{sec:pheno}, we discuss the theoretical framework relevant for our study. Section~\ref{sec:oot} contains a brief overview of OOT. In Section~\ref{sec:col}, we discuss the collider analysis for the channel considered. We present our detailed numerical analysis and results in Section~\ref{sec:result}. Finally, we summarize and conclude our discussion in Section~\ref{sec:con}.

\section{Theoretical framework}
\label{sec:pheno}
The deviation of the self-interactions of gauge bosons from the SM is considered to be one of the most important probe of new physics (NP) beyond standard model (BSM). The self-interaction of gauge bosons\footnote{interaction between charged gauge boson ($W^\pm$) with neutral gauge bosons ($\gamma, Z$)} within SM framework can be understood by the non-Abelian $SU(2)_L \times U(1)_{Y}$ gauge theory. The gauge sector 
Lagrangian involving only gauge bosons within SM is written as,
\beq
\mathcal{L}_{gauge}=-\frac{1}{4}B_{\mu \nu} B^{\mu \nu} - \frac{1}{4}W^{i}_{\mu \nu} W^{i \mu \nu}.
\label{eq:lgauge}
\eeq
Eq.~\eqref{eq:lgauge} provides the necessary gauge boson self-interactions within SM at tree level. The tree-level gauge-boson vertex can be penned down as, 
\beq
V_{W_{\nu}^+(q) W_{\lambda}^-(r) V_{\mu}(p)}=i g_{W^+W^-V}\big(g_{\mu \nu}(p-q)_{\lambda} + g_{\mu \lambda}(p-q)_{\nu} + g_{\nu \lambda}(p-q)_{\mu}\big),
\eeq
with $V=\gamma, Z$ and
\beq
g_{W^+ W^- \gamma} = g \sin \theta_w=e_0, \quad  g_{W^+ W^- Z} = g \cos \theta_w;
\eeq  
where $e_0$ is the $U(1)_{\tt em}$ coupling constant, $g$ is the $SU(2)_L$ coupling constant and $\theta_w$ is the weak-mixing angle. Therefore, we can see that there is no interaction between $\gamma$ and $Z$ at tree level in SM because $Z$ boson doesn't possess any electromagnetic charge. Therefore, the interaction between $\gamma$ and $Z$ which we call aNTGCs ($ZZ\gamma$ and $Z\gamma\gamma$), play a crucial role in investigating any NP beyond Standard Model (BSM). 

Standard model effective filed theory (SMEFT) \cite{Buchmuller:1985jz,Grzadkowski:2010es,Lehman:2014jma,Bhattacharya:2015vja,Murphy:2020rsh,Li:2020gnx} is a highly adequate framework to parameterize any small deviation from SM in a model-independent manner. The effective Lagrangian, in addition to the standard model (SM), in presence of higher dimensional operators can be written as,

\beq
\mathcal{L}=\mathcal{L}_{SM}+\sum_{d>4}\sum_{i}\frac{C_i}{\Lambda^{d-4}}\mathcal{O}^{d}_{i},
\eeq
where $\Lambda$ is the scale of new physics, $d$ is the dimension of EFT operators $\mathcal{O}_{i}^{d}$'s, which are made of SM fields and respect SM gauge symmetry. $C_i$'s are the respective Wilson coefficients which act as useful parameters to probe any NP effect. The aNTGCs are absent in dimension-6 EFT operators at the tree-level but they are present in 1-loop level. On the other hand, dimension-8 operators provide the desired aNTGCs at tree level. We will now focus on the dimension-8 operators contributing to aNTGCs. Here, we note that, apart from aNTGCs, other vertices involved in the process $e^+\,e^- \to Z \gamma$, such as $e^+e^-\gamma/Z$ can get NP contribution from dimension-6 SMEFT operators at tree level. However, from the results of electroweak precision test at LEP2 and LHC, the deviations of $Zf\bar{f}$ couplings are constrained $\lsim 0.1\%$ and that of $e^+e^-\gamma$ coupling $\ll 0.1\%$ \cite{ParticleDataGroup:2022pth}. Therefore, in this analysis, for optimal measurements of aNTGCs through $Z \gamma$ production at the $e^+e^-$ colliders, we have ignored their effect. For t-channel $W$-mediated $e^+e^- \to \nu \nu \gamma$ final state, will have negligible contribution to our signal, since we demand production of the neutrinos from on-shell $Z$. In principle, for this process, $e \nu W$ and $WW\gamma$ vertices can also get contribution from dimension-6 SMEFT operators. However, experimental measurements at LEP2 and LHC allow for only $\sim 0.1\%$ deviation for $e \nu W$  coupling from SM and for $WW\gamma$ coupling, the deviation is $\ll 0.1\%$ \cite{ParticleDataGroup:2022pth}. Therefore, the effect of dimension-6 operators contribution is also neglected in this case.

\subsection{Dimension eight effective operators}
\label{sec:dim8}

The dimension-8 effective operators that provide $ZZ\gamma$ and $Z \gamma \gamma$ couplings can be written as \cite{Degrande:2013kka}
\bea \begin{aligned}
	\mathcal{O}_{\tilde B W}&=iH^{\dagger}\tilde{B}_{\mu\nu}W^{\mu\rho}\{D_{\rho},D^{\nu}\}H,\\
	\mathcal{O}_{ B W}&=iH^{\dagger}B_{\mu\nu}W^{\mu\rho}\{D_{\rho},D^{\nu}\}H,\\
	\mathcal{O}_{ W W}&=iH^{\dagger}W_{\mu\nu}W^{\mu\rho}\{D_{\rho},D^{\nu}\}H,\\
	\mathcal{O}_{ B B}&=iH^{\dagger}B_{\mu\nu}B^{\mu\rho}\{D_{\rho},D^{\nu}\}H,\\
	\label{eq:dim8}
\end{aligned}
\eea
where $B_{\mu\nu}$ and $W^{\mu\nu}$ are the gauge field strength tensors and $D_{\mu}$ is the covariant derivative. In Eq.~\eqref{eq:dim8}, $\mathcal{O}_{\tilde B W}$ is CP-conserving operator and the rest of three are CP-violating operators. The definitions of field strength tensors and covariant derivative are as follows:
\begin{align}
	B_{\mu \nu}&= (\partial_{\mu} B_{\nu}-\partial_{\nu} B_{\mu}),\\
	W_{\mu \nu}&=\sigma^{i} (\partial_{\mu} W^{i}_{\nu}-\partial_{\nu} W^{i}_{\mu}+g \epsilon_{ijk}W_{\mu}^{j}W_{\nu}^{k}),\\
	D_{\mu}&=\partial_{\mu}-ig W_{\mu}^{i} \sigma^{i}-i\frac{g^{'}}{2}B_{\mu}Y.
\end{align}

ATLAS experiments put the latest experimental bound on the dimension-8 couplings through $pp \to Z\gamma\to\nu \bar \nu \gamma$ channel at center-of-mass energy ($\sqrt{s}$) of 13 TeV with integrated luminosity of 36.1 $\rm fb^{-1}$ at LHC \cite{ATLAS:2018nci}. At 95\% C.L., the experimental bounds on the aNTGCs without any systematic errors are given as,
\bea\begin{aligned}
	-1.1 \, \rm{TeV^{-4}} < \frac{C_{\tilde B W}}{\Lambda^4} < 1.1 \, \rm{TeV^{-4}},\\
	-2.3 \, \rm{TeV^{-4}} < \frac{C_{W W}}{\Lambda^4} < 2.3 \, \rm{TeV^{-4}},\\
	-0.65 \, \rm{TeV^{-4}} < \frac{C_{ B W}}{\Lambda^4} < 0.64 \, \rm{TeV^{-4}},\\
	-0.24 \, \rm{TeV^{-4}} < \frac{C_{B B}}{\Lambda^4} < 0.24 \, \rm{TeV^{-4}}.\\
\end{aligned} 
\eea
	
	\noindent
	As we discussed earlier, the dimension-6 operators do not give rise to any aNTGCs at tree level but their contribution become important at one-loop level. However, the order of the contribution from one-loop level in case of dimension-6 operators are roughly $\mathcal{O}(\frac{\alpha_{\tt EM} s}{4 \pi \Lambda^2})$, whereas for dimension-8 operators the tree-level contribution becomes $\mathcal{O}(\frac{s v^2}{\Lambda^4})$. Evidently, the contribution of the dimension-8 operators to the aNTGCs supersedes the contribution of dimension-6 operators in the limit of $\Lambda < \sqrt{\frac{4 \pi}{\alpha_{\tt EM}}}v\sim 10$ TeV~\cite{Degrande:2013kka}. However, it should be carefully noted that, if the loop-induced dimension-6 contribution involves electroweak coupling instead of electromagnetic coupling (depending on the particular loop diagram in consideration), cut-off on $\Lambda$ comes down to $\sim 5$ TeV. The aforementioned limits are derived with the assumption that the Wilson coefficients pertaining to the dimension-6 and -8 operators are roughly equal.
	
	The resulting effective Lagrangian that contains aNTGCs from dimension-6 and dimension-8 operators is given by
		\begin{align}
		\nonumber
		\mathcal{L}_{\tt EFT}=&\frac{g_e}{m_Z^2}\bigg[-\big\{f_{4}^{\gamma}(\partial_{\mu}F^{\mu \nu})+f_{4}^{Z}(\partial_{\mu}F^{\mu \beta})\big\}Z_{\alpha} (\partial^{\alpha} Z_{\beta})+\big\{f_{5}^{\gamma}(\partial^{\sigma}F_{\sigma \mu})+f_{5}^{Z}(\partial^{\sigma}Z_{\sigma \mu})\big\}\tilde{Z}^{\mu \beta} Z_{\beta}\\\nonumber
		&-\big\{h_{1}^{\gamma}(\partial^{\sigma}F_{\sigma \mu})+h_{1}^{Z}(\partial^{\sigma}Z_{\sigma \mu})\big\} Z_{\beta} F^{\mu \beta}-\big\{h_{3}^{\gamma}(\partial_{\sigma}F^{\sigma \rho})+h_{3}^{Z}(\partial_{\sigma}Z^{\sigma \rho})\big\} Z^{\alpha} \tilde{F}_{\rho \alpha}\\\nonumber
		&-\bigg\{\frac{h^{\gamma}_2}{m^2_Z}(\partial_{\alpha} \partial_{\beta} \partial^{\rho} F_{\rho \mu}) +\frac{h^{Z}_2}{m^2_Z} \big(\partial_{\alpha} \partial_{\beta}(\Box + m_Z^2)Z_{\mu}\big)\bigg\}Z^{\alpha} F^{\mu \beta}-\bigg\{\frac{h^{\gamma}_4}{2m^2_Z}(\Box \partial^{\alpha} F^{\rho \alpha})\\
		& +\frac{h^{Z}_4}{2m^2_Z} \big((\Box + m_Z^2) \partial^{\sigma} Z^{\rho \alpha}\big)\bigg\}Z_{\alpha} \tilde{F}_{\mu \beta}\bigg],
	\end{align}
	
	\noindent
	where $\tilde{Z}_{\mu \nu}=\frac{1}{2}(\epsilon_{\mu \nu \rho \sigma} Z^{\rho \sigma})$ and $Z_{\mu \nu}=(\partial_{\mu} Z_{\nu}-\partial_{\nu} Z_{\mu})$ are the field strength tensor. Here, $f_3^{V}, \, f_4^{V}, \, f_5^{V}$ ($V=\gamma, Z$) are dimension-6 couplings and $h_3^{V}, \, h_4^{V}, \, h_5^{V}$ are the dimension-8 couplings. The expressions of aNTGCs in terms CP conserving dimension-8 coupling are written as;
	
	\begin{align}
		h_3^Z&=\frac{v^2 m_Z^2 C_{\tilde{B}W}}{4 c_w s_w \Lambda^4},\\
		h_4^Z&=h_3^{\gamma}=h_4^{\gamma}=0,\\\nonumber
	\end{align}
	
	whereas, for CP-violating case, the couplings can be written as,
	
	\begin{align}
		h_1^Z&=\frac{m_Z^2 v^2\left(-c_w s_w C_{WW}+C_{BW}(c_w^2-s_w^2)+4 c_w s_w C_{BB}\right)}{4 c_w s_w \Lambda^4}, \\
		h_1^{\gamma}&=\frac{m_Z^2 v^2\left(s_w^2 C_{WW}-2 c_w s_wC_{BW}+4 c_w^2 C_{BB}\right)}{4 c_w s_w \Lambda^4},\\
                h_2^\gamma&=h_2^Z=0. \\
	\end{align}
	
	\noindent
	In the subsequent analysis, we will be discussing the statistical limit on dimension-8 aNTGCs considering the highest center of mass energy ($\sqrt{s}$) and maximum polarization combination of the incoming beams with same luminosity in the context of the future linear colliders such as ILC  and CLIC. The design details are presented in table~\ref{tab:design}.
	\begin{center}
		\begin{table}[htb!]
			\begin{tabular}{ |c |c |c |c | c| c| }
				\hline
				\multicolumn{1}{|c|}{Linear} &
				\multicolumn{1}{|c|}{ c.o.m energy ($\sqrt{s}$)} &
				\multicolumn{1}{|c|}{luminosity ($\mathcal{L}_{int}$)} &
				\multicolumn{3}{ c|}{beam} \\ 
				\multicolumn{1}{|c|}{colliders} &
				\multicolumn{1}{c|}{(TeV)} &
				\multicolumn{1}{c|}{($\rm fb^{-1}$)} &
				\multicolumn{1}{c}{} & 
				\multicolumn{1}{c}{polarization} &
				\multicolumn{1}{c|}{} \\
				\cline{4-6}
				\hline
				ILC & 1 & 1000 & $P_{e^\pm} = ^{00\%}_{00\%}$ & $P_{e^\pm} = ^{-30\%}_{+80\%}$ & $P_{e^\pm} = ^{+30\%}_{-80\%}$ \\ 
				\hline
				CLIC & 3 & 1000 & $P_{e^\pm} = ^{00\%}_{00\%}$ & $P_{e^\pm} = ^{+00\%}_{+80\%}$ & $P_{e^\pm} = ^{+00\%}_{-80\%}$ \\ 
				\hline
			\end{tabular}
			\caption{Design details of ILC and CLIC.}
			\label{tab:design}
		\end{table}
	\end{center}

\section{Optimal Observable Technique}
\label{sec:oot}

The optimal observable technique (OOT) is an effective tool to estimate the precision of NP coupling measurement in an economical way. Here, we briefly outline the mathematical framework of OOT which has already been discussed in \cite{Diehl:1993br,Gunion:1996vv} in detail. In general, a collider observable ({\it e.g.} differential cross section) containing the contribution from the SM and BSM can be written in the form 
\beq
\mathcal{O}(\phi)=\frac{d\sigma_{\tt theo}}{d\phi} = \sum_i g_i f_i(\phi) \,,
\label{eq:expnd1}
\eeq
where $ \phi $ indicates some suitable phase-space variable, the coefficients $g_i $ are the functions of NP couplings and numerical constants, and  $ f_i $'s are  linearly-independent functions of the phase space variable $\phi$. In the following, as we will explore 2 $\rightarrow$ 2 scattering process ($e^+e^-\rightarrow Z\gamma$), the cosine of the scattering angle of $\gamma$ ($\cos\theta$) is the phase-space variable in consideration. In principle, $\phi$ can be chosen any other observable as well, depending on the process.

Now,  we consider a realistic experimental scenario where the event rate is constant over a finite time, and the event number follows a Poisson distribution. The primary focus is to determine $g_i$'s. Therefore, by using suitable weighting function ($w_i(\phi)$), $g_i$'s can be estimated as:

\beq
g_i=\int w_i(\phi) \mathcal{O}(\phi) d\phi.
\eeq

In general, several choices of $w_i(\phi)$ are possible, but there is a unique choice for which the covariance matrix ($V_{ij}$) is optimal in a sense that the the statistical uncertainties in $g_i$'s are minimized. For this choice, $V_{ij}$ is presented as;

\beq
V_{ij} \propto \int w_i({\phi})w_j({\phi}) \mathcal{O}({\phi}) d\phi.
\eeq

Therefore, the weighting functions subject to the condition $\delta V_{ij}=0$  are 

\beq
w_i(\phi)=\frac{M_{ij}^{-1}f_j(\phi)}{\mathcal{O}(\phi)},
\eeq

where,

\beq
M_{ij} =\int \frac{f_i(\phi) f_j(\phi)}{\mathcal{O}(\phi)}d\phi.
\label{eq:mij}
\eeq

Then, the optimal covariance matrix becomes 
\beq
V_{ij} =\frac{ M_{ij}^{-1} \sigma_T}{N}= \frac{M^{-1}_{ij}}{\mathfrak{L}_{\tt int}}\, ,
\label{eq:covmat1}
\eeq
with
where $\sigma_T=\int \mathcal{O}(\phi) d\phi$ and N is total number of events ($N=\sigma_T \mathfrak{L}_{\tt int}$). $\mathfrak{L}_{\tt int}$ denotes the integrated luminosity over this period.

The $\chi^2$ function that measures the accuracy of NP couplings is defined as 

\beq
\chi^2= \sum_{\{i,j\}=1}^{n} (g_i -g_i^0) (g_j -g_j^0) \,  \left( V^{-1}\right)_{ij},
\label{eq:chi2}
\eeq
where, $g^0$'s are the {\it `seed values'} that depend on the NP model. The limit dictated by $\chi^2 \le n^2$ corresponds to $n\sigma$ standard deviation from a seed values ($g^0$)  is the optimal limit for any NP couplings as the covariance matrix ($V_{ij}$) is minimal. Using the definition of $\chi^2$ functions in Eq.~\eqref{eq:chi2}, the optimal limits on the NP couplings has been discussed in the following sections.

\section{Collider simulation}
\label{sec:col}

In order to probe the aNTGCs at the colliders, one has to produce copious number of signal events. At the same time, the signal events have to be significant over and above the non-interfering SM backgrounds in order to make precision measurements of the anomalous coupling. With this in view, we perform a collider analysis and try to obtain suitable conditions for most precise estimation of aNTGCs. 
In our analysis, we will consider the signal process ($e^+e^- \rightarrow Z{(\nu \bar{\nu}) \gamma}$). This final state has various advantages over the processes where $Z$ decays into charged leptons or hadronic final states. In case of $Z$ decaying to hadrons, the final state will contain large multi-jet background. On the other hand, a larger $Z$ boson branching ratio into neutrinos compared to that into charged leptons provides an opportunity to study the $Z\gamma$ production in high $p_T$ region, where the sensitivity of the anomalous couplings will be higher. 

\begin{figure}[htb!]
	$$
	\includegraphics[height=4.8cm, width=6.8cm]{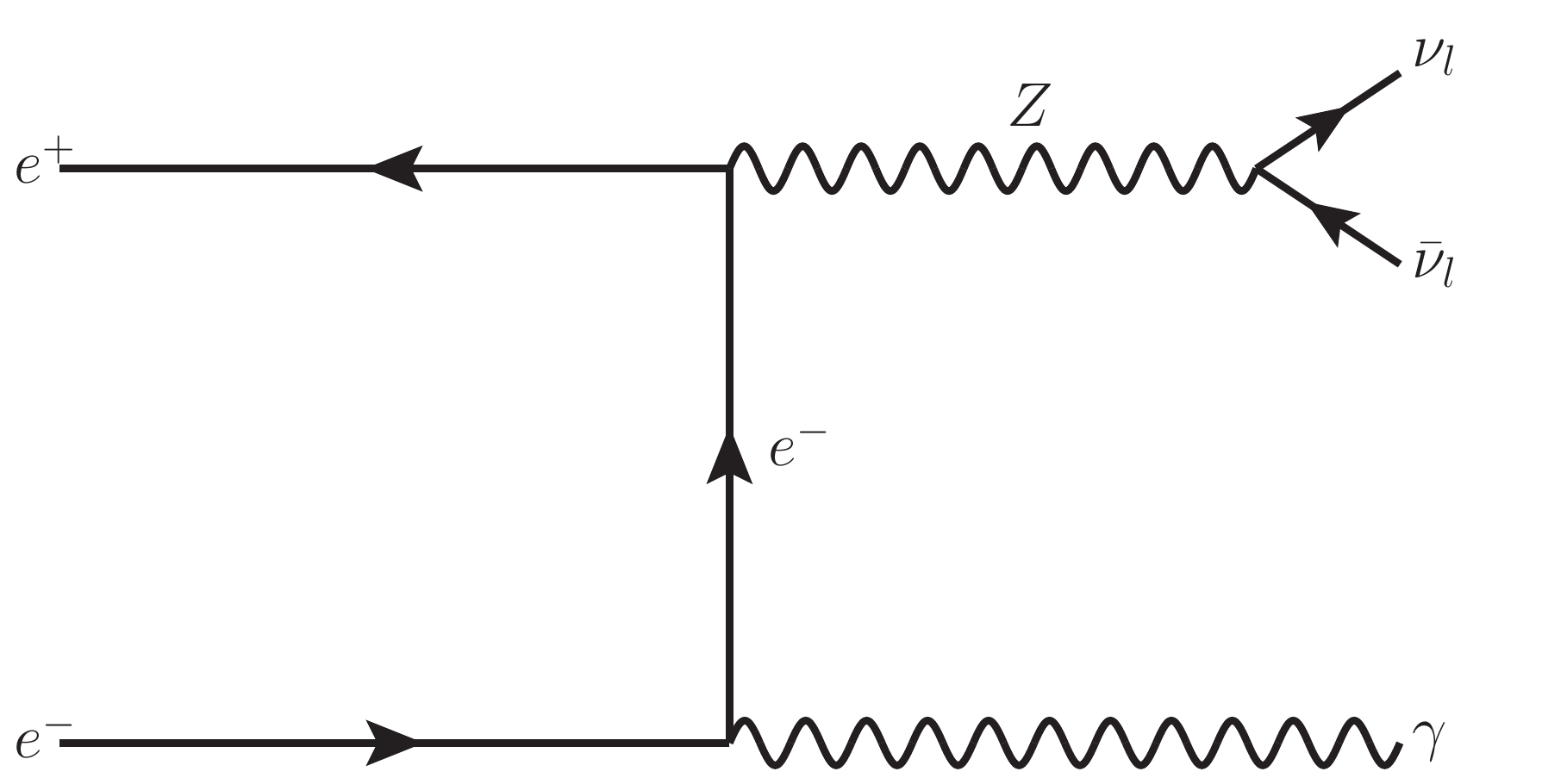}
	\includegraphics[height=4.8cm, width=6.8cm]{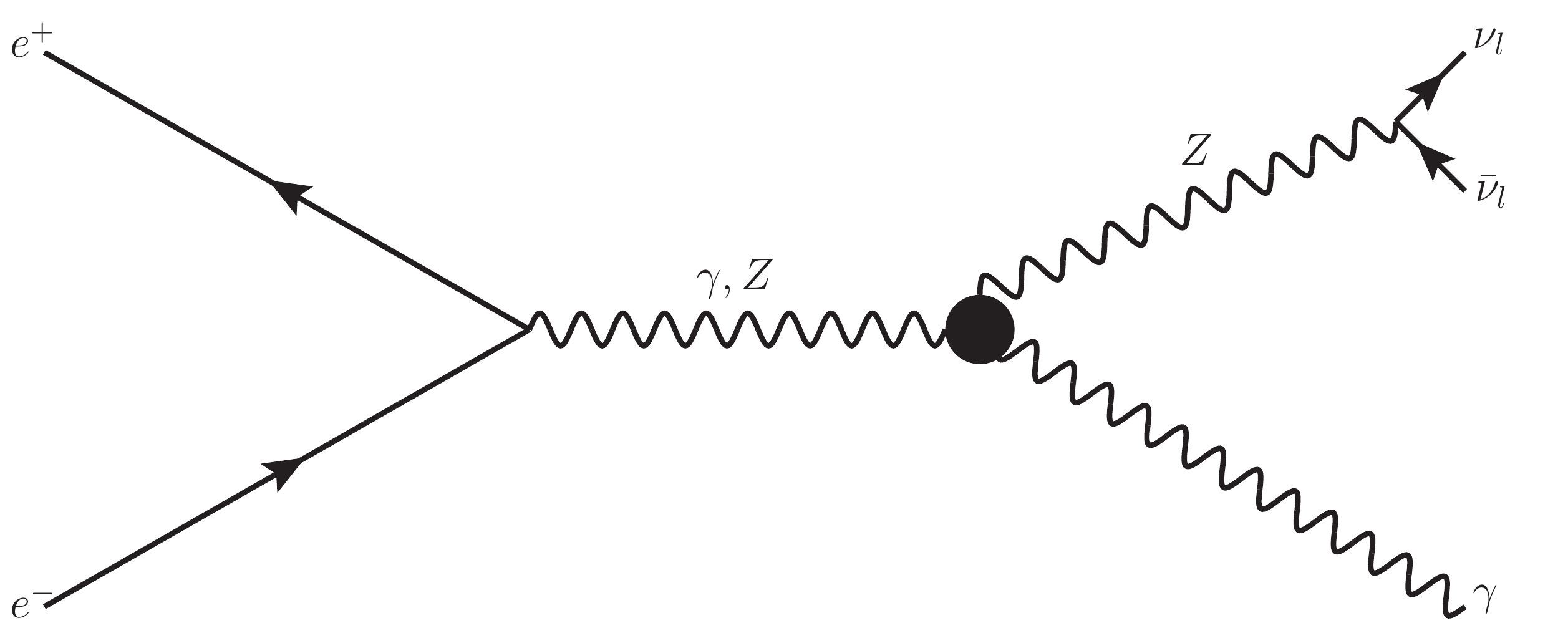}
	$$
	\caption{Production and decay of $Z\gamma$ final state at $e^+\,e^-$ colliders; left: SM contribution; right: SMEFT contribution.}
	\label{fig:tcprod}
\end{figure}

\noindent
Therefore the signal of our interest is $e^+e^- \rightarrow \nu\nu\gamma$ (See figure~\ref{fig:tcprod}). We demand exactly one photon with $p_T > 10$ GeV and $|\eta| < 2.5$. In addition, veto on jets with $p_T > 20$ GeV and leptons with $p_T > 10$ GeV is also imposed in the final state. The major background process which gives rise to the same final state is SM $\nu\bar\nu\gamma$ where the photon is emitted off a $t$-channel $W$ boson. This background is irreducible. There are also other backgrounds like $t\bar t \gamma$, $W^+W^-\gamma$, which can contribute to our desired final state when the hadronic/leptonic the decay products from top or $W$'s are very soft and therefore escape the detector. 
The signal and background events are generated in {\tt Madgraph5@NLO}~\cite{Alwall:2014hca}.
Detector simulation is taken care of by {\tt Delphes}(v3)~\cite{deFavereau:2013fsa}. The UFO file for the NP model has been generated via Feynrules~\cite{Christensen:2008py,Degrande:2013kka}.
We have plotted two crucial kinematical observables missing energy ($\slashed{E}$) and missing transverse momenta ($\slashed{E_T}$) in order to distinguish signal from the backgrounds. The observables are defined as follows.

\begin{itemize}
	\item {\it Missing Transverse Energy or $\slashed{E}_T$:} The vector sum of transverse momenta of all the missing particles (not registered in the detector) can be estimated from the momentum imbalance in the transverse direction associated with all visible particles. Thus $\slashed{E_T}$ is defined as:
	\bea
	\slashed{E}_T = -\sqrt{(\sum_{\ell,j} p_x)^2+(\sum_{\ell,j} p_y)^2},
	\eea
	where the sum runs over all the visible objects including leptons, jets etc. 
	
	\item{\it Missing Energy or ME ($\slashed{E}$)}: The energy that is carried away by the missing final state particles, can be identified at the lepton collider given the knowledge of centre of mass energy as 
	\bea
	\slashed{E}=\sqrt{s}-\sum_{\ell,j,\gamma} E; 
	\eea
\end{itemize}

\begin{figure}[htb!]
	$$
	\includegraphics[height=7cm, width=8cm]{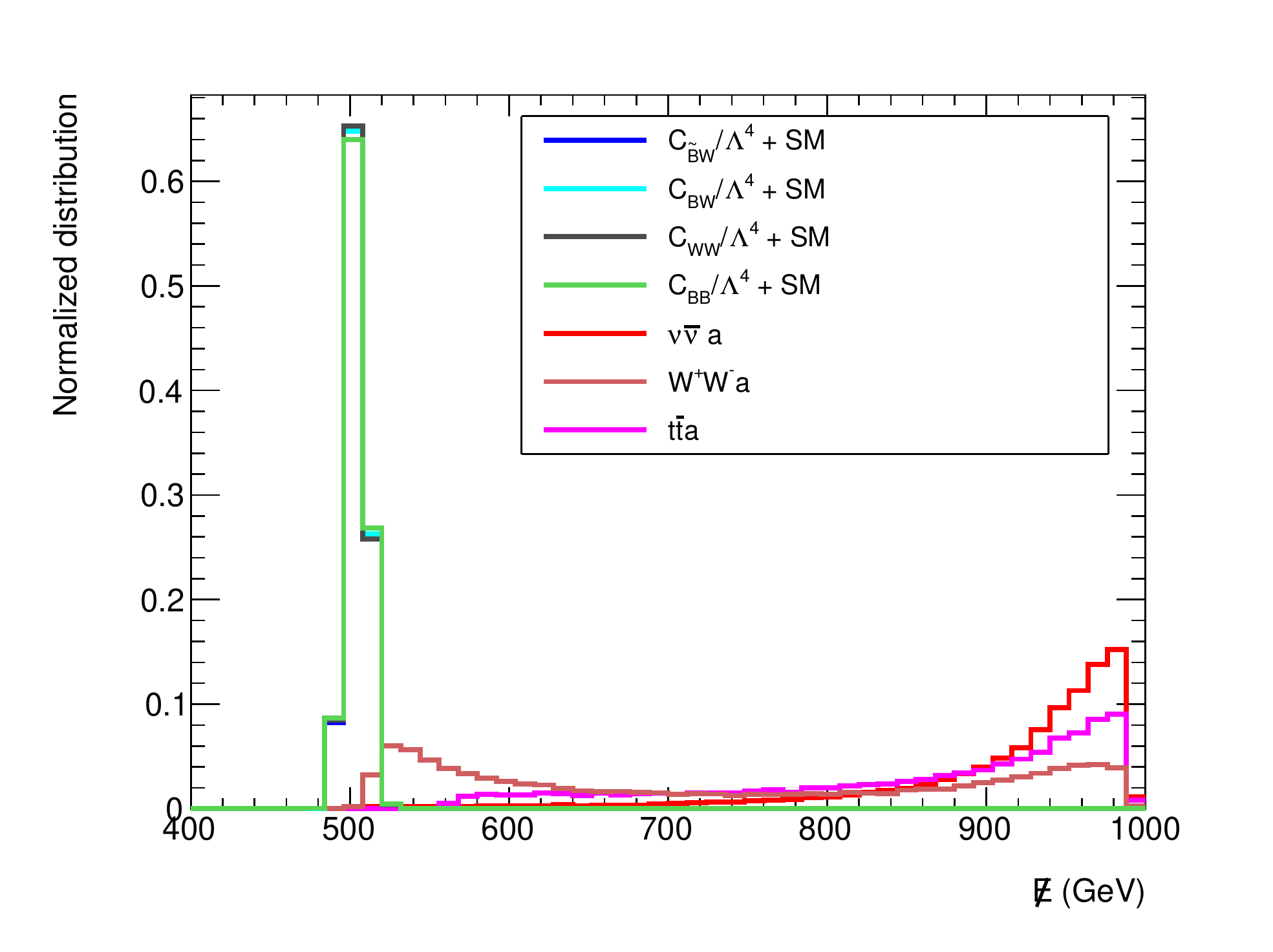}
	\includegraphics[height=7cm, width=8cm]{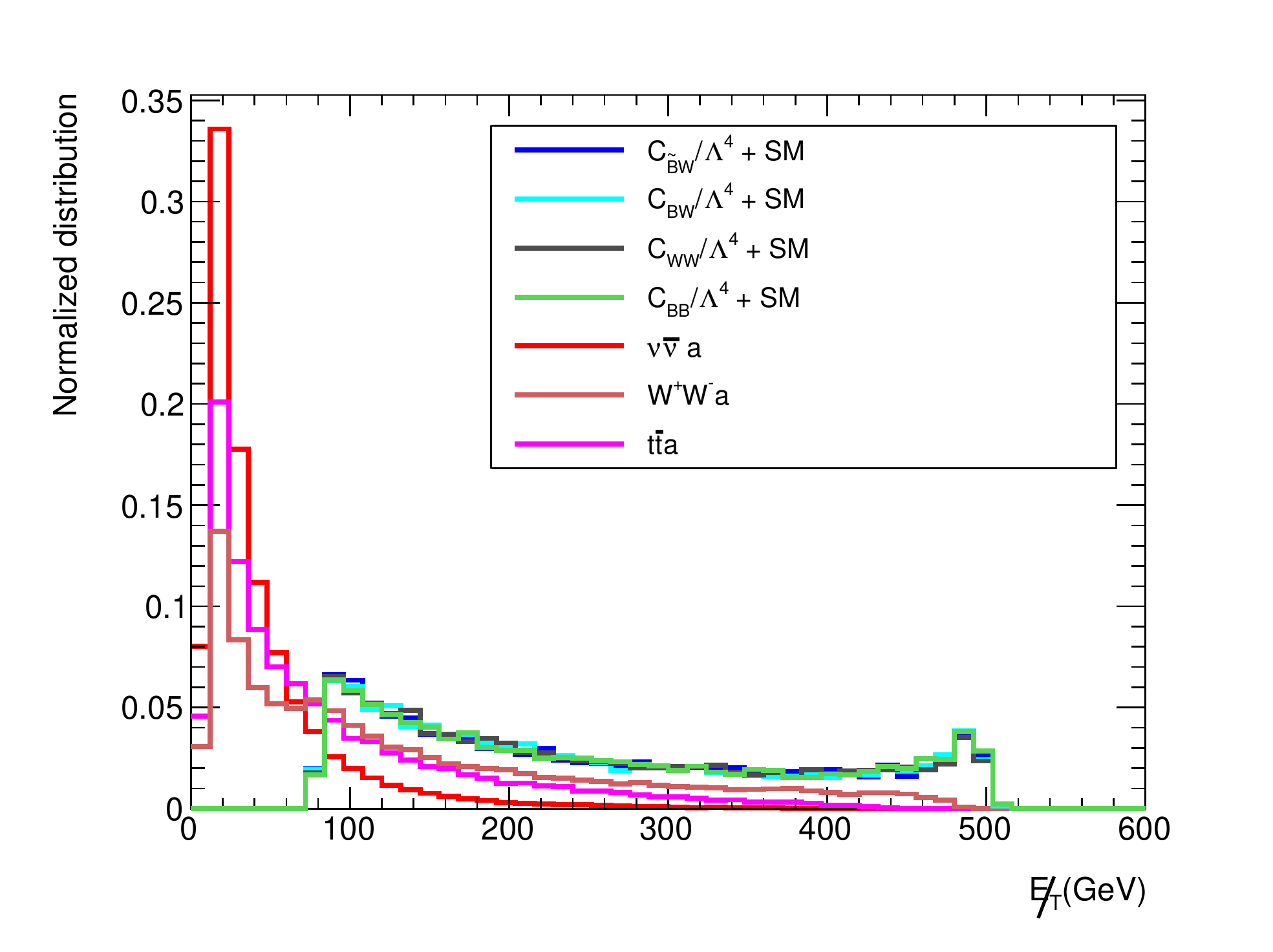}
	$$
	$$
	\includegraphics[height=7cm, width=8cm]{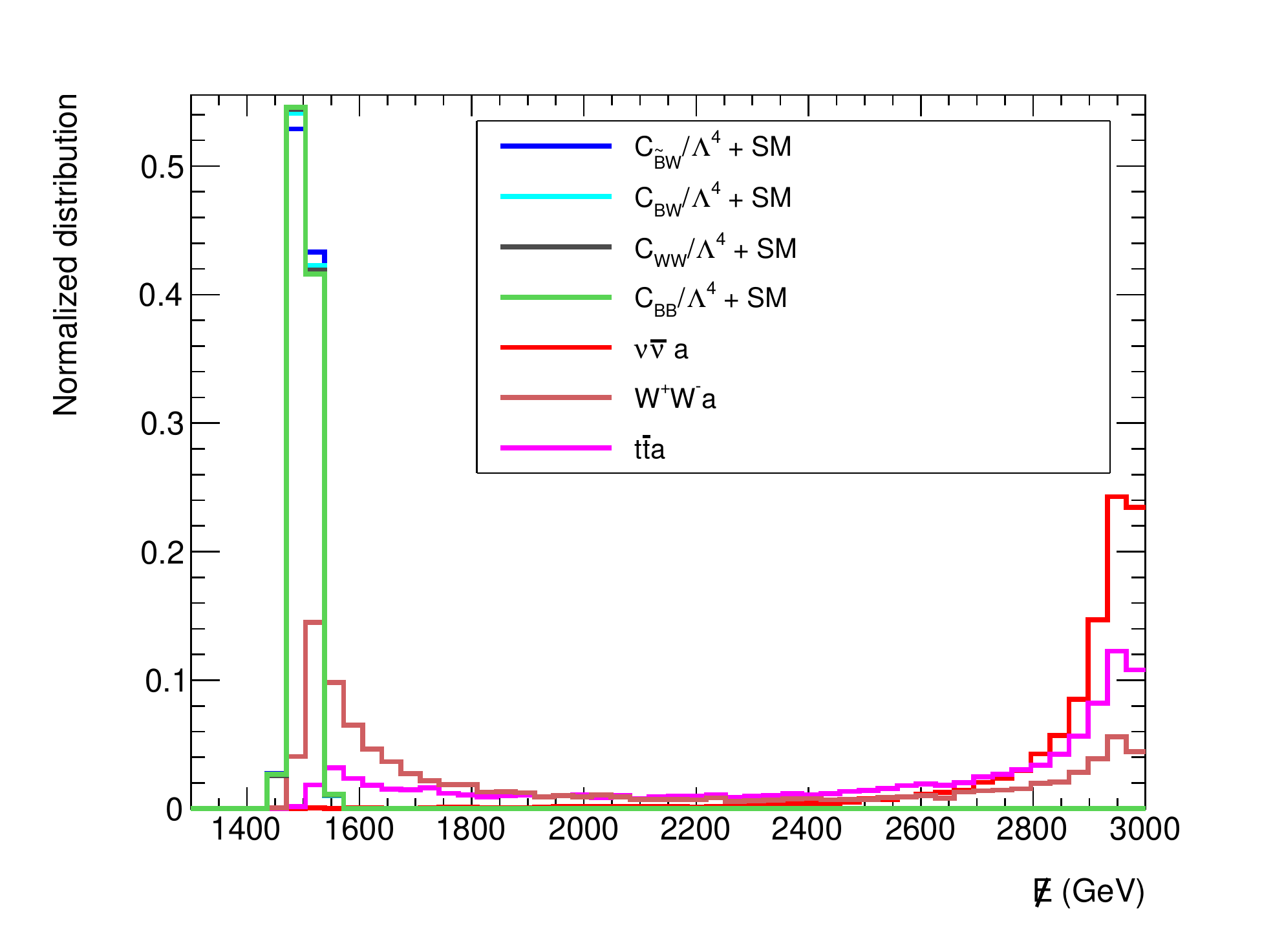}
	\includegraphics[height=7cm, width=8cm]{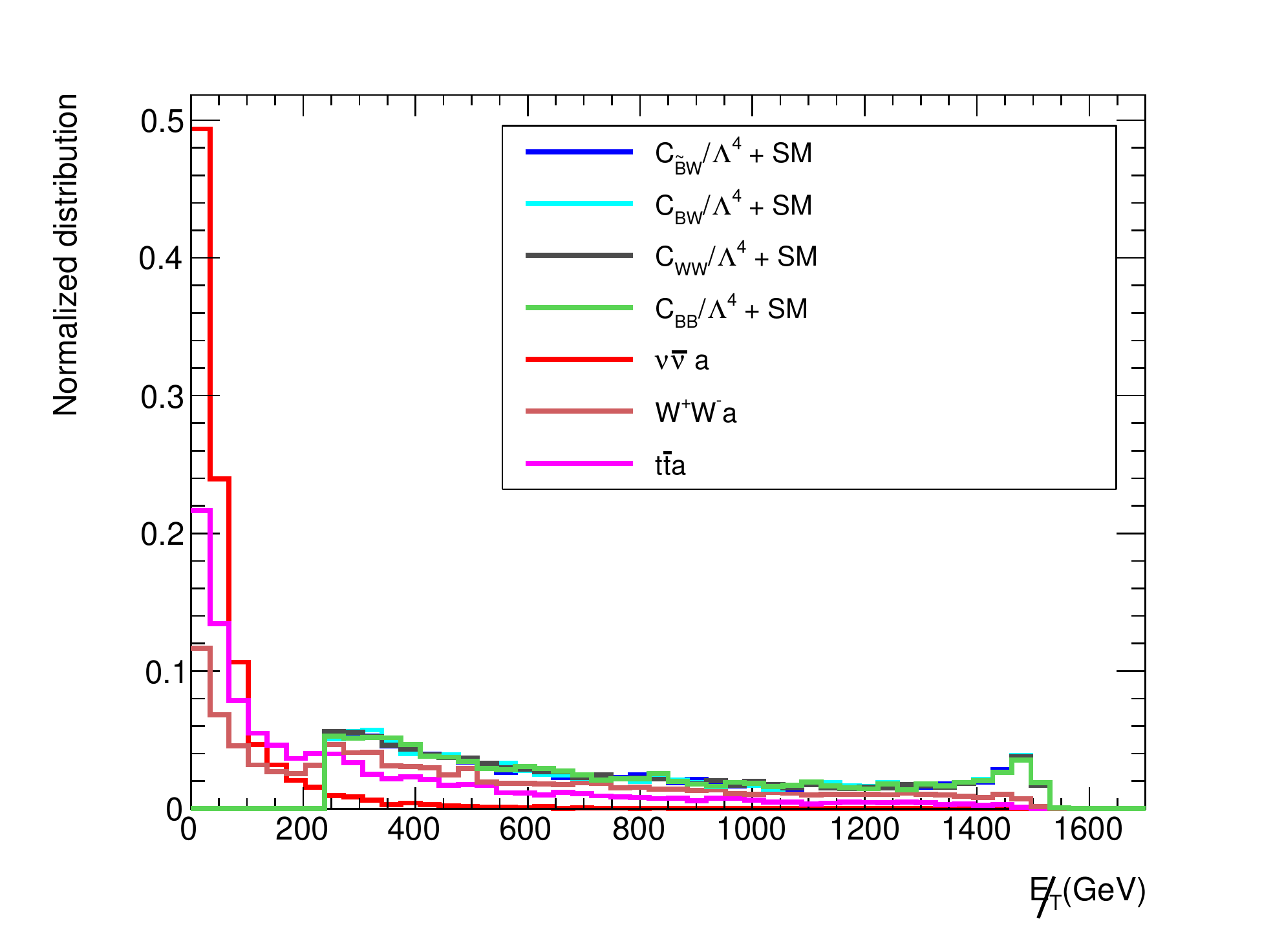}
	$$
	\caption{Normalized distribution for $e^+ e^- \rightarrow Z(\nu\bar\nu)\gamma$ (SM + aNTGCs) and non-interfering SM backgrounds with (top panel) $\sqrt{s}= 1$ TeV (ILC), $\Lambda = 1.3$ TeV, $C_{ij}=1$ and (bottom panel) $\sqrt{s}= 3$ TeV (CLIC), $\Lambda = 3.2$ TeV, $C_{ij}=1$ with unpolarized beams. Left: missing energy ($\slashed{E}$), right: missing transverse energy ($\slashed{E_T}$).} 	
	\label{fig:nodist}
\end{figure}

\noindent
In figure~\ref{fig:nodist}, we show the kinematical distributions for signal and backgrounds. For signal, we have considered CP-conserving coupling $C_{\tilde B W}$ as well as CP-violating couplings $C_{BW}$, $C_{WW}$ and $C_{BB}$ in the plots. The signal here includes the interference between SM and aNTGC couplings. Amongst the non-interfering backgrounds, the largest cross-section pertains to $\nu\bar\nu\gamma$(from $t$-channel $W$). However, we can see that though this background is irreducible, it gives rise to a three-body final state as opposed to the signal, which is a two-body final state. Therefore, the $\slashed{E}$ distribution becomes a narrow peak around $\frac{\sqrt{s}}{2}$ in case of signal, unlike the $\nu\bar\nu\gamma$ background. Therefore, a cut on $\slashed{E}$ reduces this background significantly. On the other hand, $t\bar t\gamma$ and $WW\gamma$ backgrounds are suppressed by the lepton and jet-veto. In addition, a moderate $\slashed{E_T}$ cut improves the signal significance even further. We have found that, with 1440 GeV $< \slashed{E} < 1560$ GeV and $\slashed{E_T} > 500$ GeV for CLIC ($\sqrt{s} = 3$ TeV, $\Lambda = 3.2$ TeV, $C_{\tilde B W} = 1$) and 450 GeV $< \slashed{E} < 560$ GeV and $\slashed{E_T} > 80$ GeV for ILC ($\sqrt{s} = 1$ TeV, $\Lambda = 1.3$ TeV, $C_{\tilde B W} = 1$), the signal dominates strongly over the non-interfering backgrounds. These regions are also very sensitive to the NP couplings. Therefore, in our respective analyses with OOT, we will focus on this region of phase-space and use these aforementioned cut values. We mention here that, the NP scale $\Lambda$ has been chosen to be beyond the reach of the collider experiments in case of CLIC and ILC in order to make the EFT valid. 

\section{Results}
\label{sec:result}

\begin{figure}[htb!]
	\begin{align*}
		\includegraphics[height=3cm, width=5.2cm]{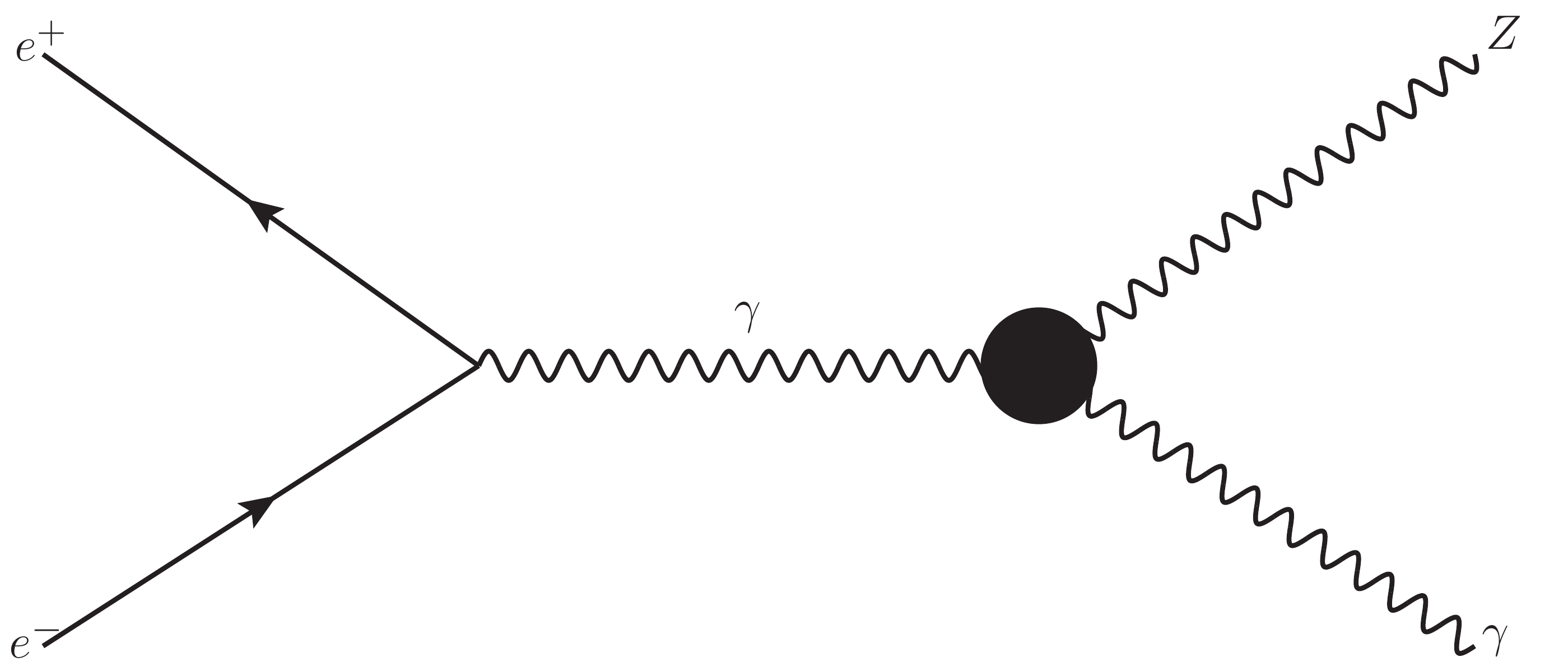} \quad&\quad
		\includegraphics[height=3cm, width=5.2cm]{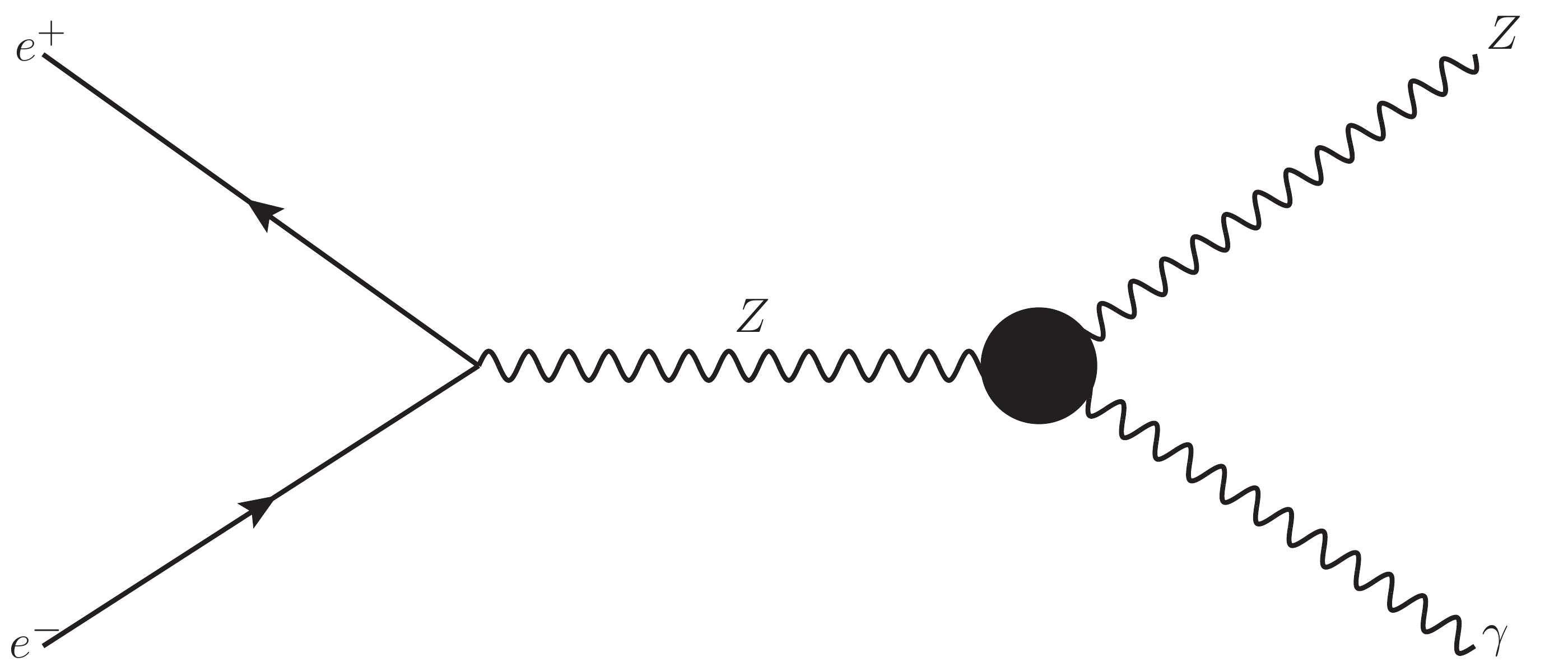} \cr
		\includegraphics[height=3cm, width=5.2cm]{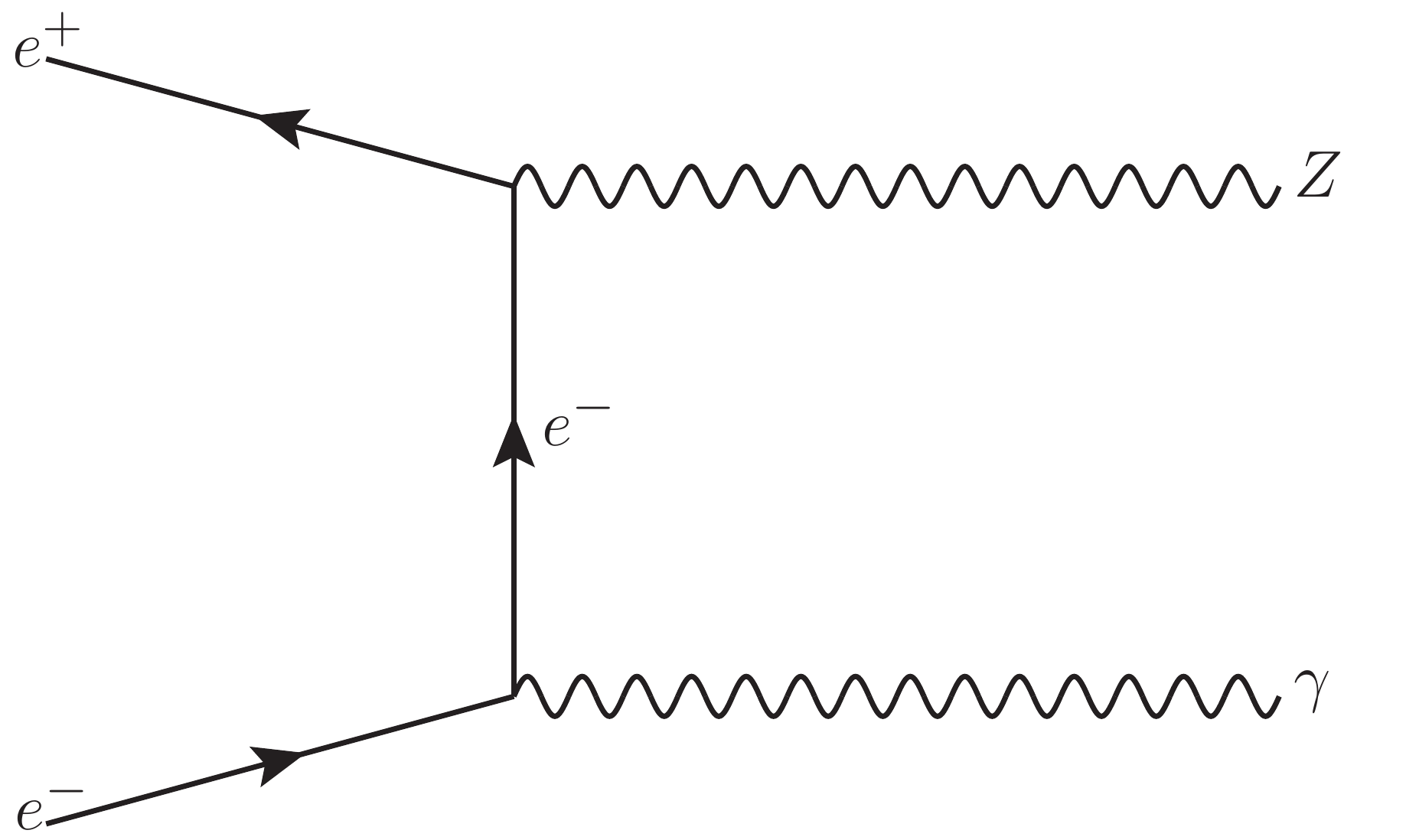} \quad&\quad
		\includegraphics[height=3cm, width=5.2cm]{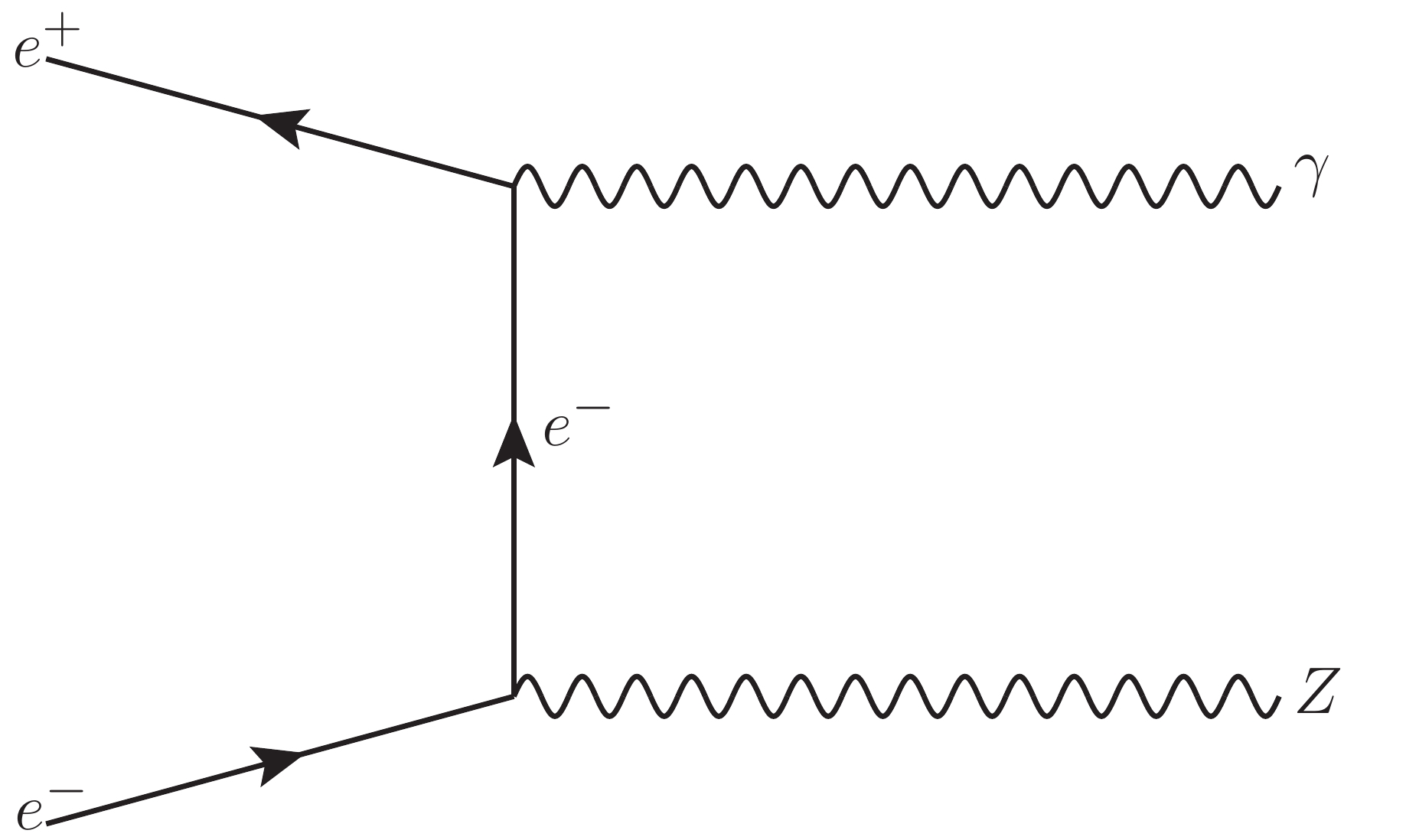} \cr
	\end{align*}
	\caption{$Z\gamma$ production at the $e^+ \, e^-$ colliders. Top panel: $ZZ\gamma$ and $Z\gamma \gamma$ contributions from dimension-8 effective operators; Bottom panel: SM contribution.} 	
	\label{fig:fynmndiag}
\end{figure}

$Z\gamma$ production at the $e^+ \, e^-$ colliders primarily takes place via the t- and u-channel diagrams within SM as shown in the bottom panel of figure~\ref{fig:fynmndiag}. The BSM contributions to the $Z \gamma$ production via dimension-8 operators written in Eq.~\eqref{eq:dim8}, lead to the s-channel diagram shown in the top panel of figure~\ref{fig:fynmndiag}. When the initial $e^\pm$ beams have partial polarizations $P_{e^\pm} $ (with $ -1\le P_{e^\pm}\le1$) the total differential cross-section is given by

\begin{align}
	\nonumber
	\frac{d\sigma(P_{e^+},\,P_{e^-})}{d\Omega} =& \frac{(1-P_{e^-})(1-P_{e^+})}4 \left( \frac{d\sigma}{d\Omega}\right)_{LL} +  \frac{(1-P_{e^-})(1+P_{e^+})}4 \left( \frac{d\sigma}{d\Omega}\right)_{LR} \\
	&\frac{(1+P_{e^-})(1-P_{e^+})}4 \left( \frac{d\sigma}{d\Omega}\right)_{RL} + \frac{(1+P_{e^-})(1+P_{e^+})}4 \left( \frac{d\sigma}{d\Omega}\right)_{RR},
	\label{eq:difcs}
\end{align}

\begin{figure}[htb!]
	\begin{align*}
		\includegraphics[height=5cm, width=4.8cm]{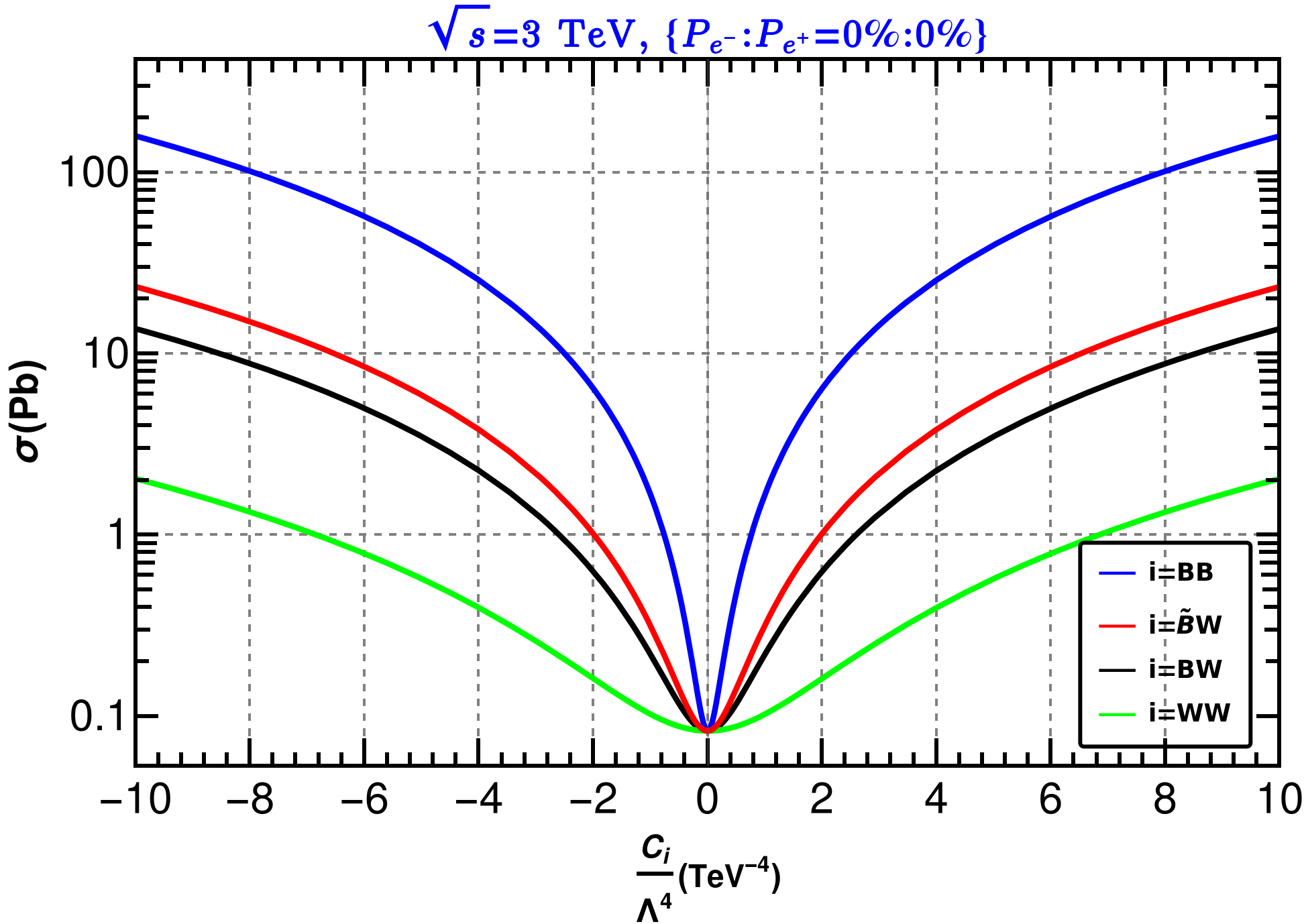} \quad
		\includegraphics[height=5cm, width=4.8cm]{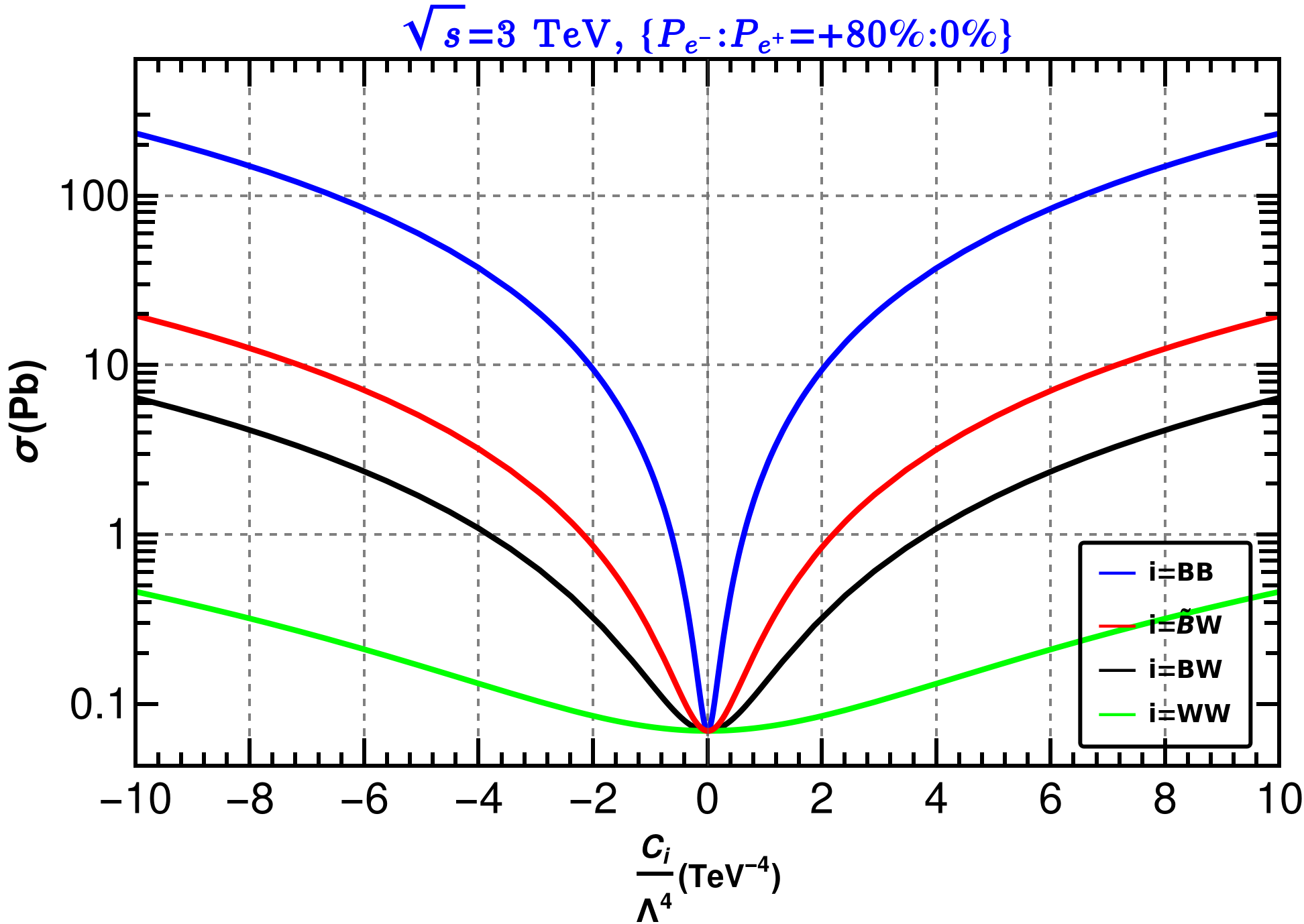} \quad
		\includegraphics[height=5cm, width=4.8cm]{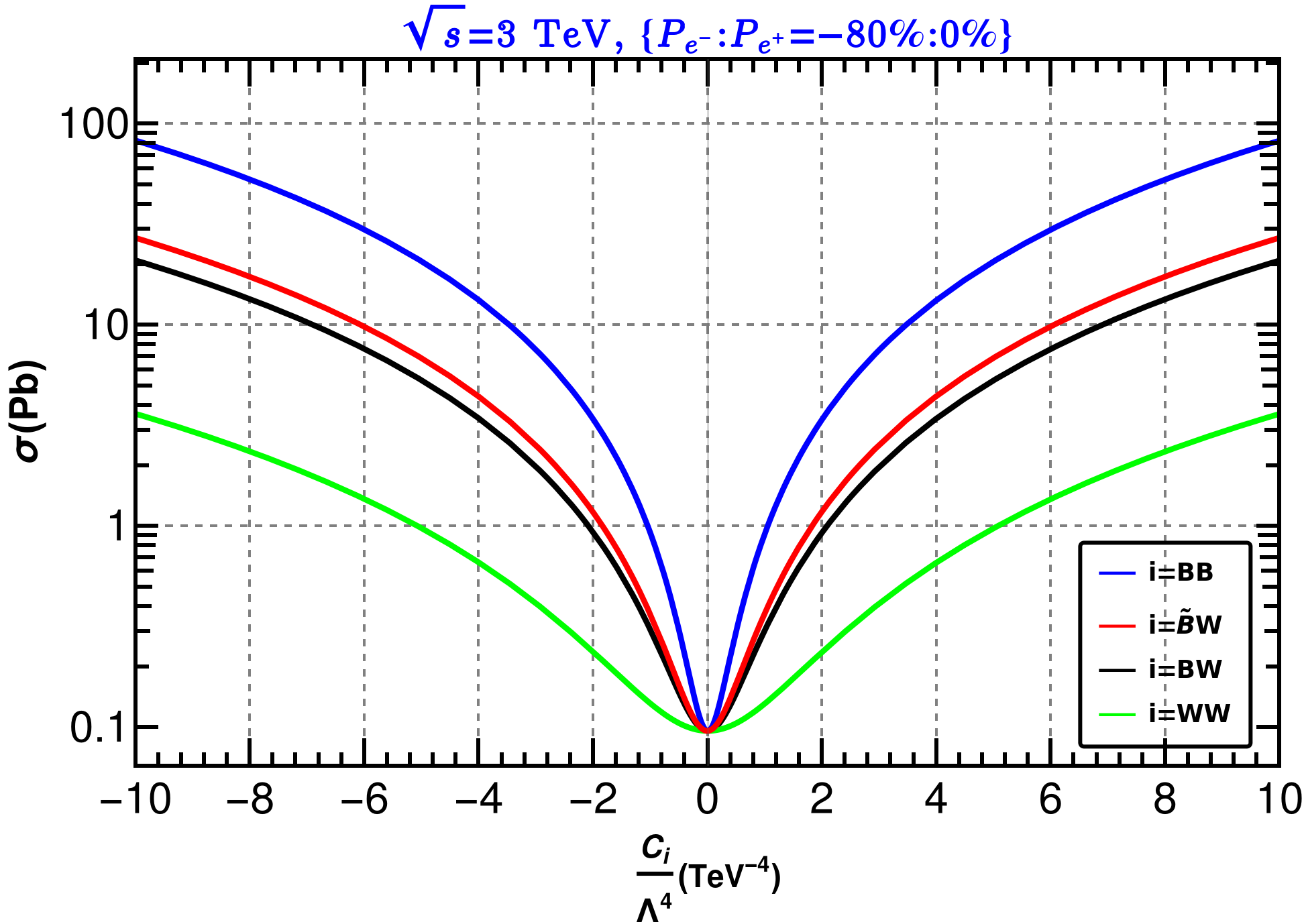} \quad
	\end{align*}
	\caption{ Variation of $Z\gamma$ production cross-section with the different dimension-8 couplings at the $e^+ \, e^-$ colliders with $\sqrt s$ = 3 TeV.  Left: $\{P_{e^-}:P_{e^+}=0\%:0\%\}$; Middle: $\{P_{e^-}:P_{e^+}=+80\%:0\%\}$; Right:$\{P_{e^-}:P_{e^+}=-80\%:0\%\}$.} 	
	\label{fig:xsecunpol}
\end{figure}

\noindent
where, $\left(\frac{d\sigma}{d\Omega}\right)_{ij}$ are the helicity amplitudes i.e. the differential cross section when the electron and positron have `$i (=L,R)$' and `$j(=L,R)$' helicities. `$L$' and `$R$' denote the left and right-handed initial beams respectively. The variation of total cross section with the NP couplings is shown in the figure \ref{fig:xsecunpol} for unpolarized beams as well as for two different polarization combinations. Total cross-section has one pure SM term, an interference term to between SM and dimension-8 operators $\left(\frac{C_i}{\Lambda^4}\right)$ and contribution solely from dimension-8 operators proportional to $\left(\frac{C_i}{\Lambda^4}\right)^2$. For both unpolarized and polarized beams, the dominant BSM contribution to the cross-section will come from the aforementioned interference term as long as the the NP couplings are small, so the variation of cross-section with the NP couplings will not be symmetric in this region. When NP couplings increase further, the NP(squared amplitude) term dominates in the total cross-section. In this region, the cross-section behaves symmetrically with the NP couplings as can be seen in figure~\ref{fig:xsecunpol}. For $\left(\frac{C_{\tilde B W}}{\Lambda^4}\right)$, $\left(\frac{C_{BW}}{\Lambda^4}\right)$ and $\left(\frac{C_{WW}}{\Lambda^4}\right)$, $\{P_{e^-}:P_{e^+}=-80\%:0\%\}$ beam polarization combination provides increment over the unpolarized cross-section whereas, for $\left(\frac{C_{BB}}{\Lambda^4}\right)$ coupling, cross-section increases for the polarization combination $\{P_{e^-}:P_{e^+}=+80\%:0\%\}$.

\subsection{Sensitivity of NP couplings}
\label{sec:sensitivity}
\begin{figure}[htb!]
	$$
	\includegraphics[height=6.8cm, width=7.6cm]{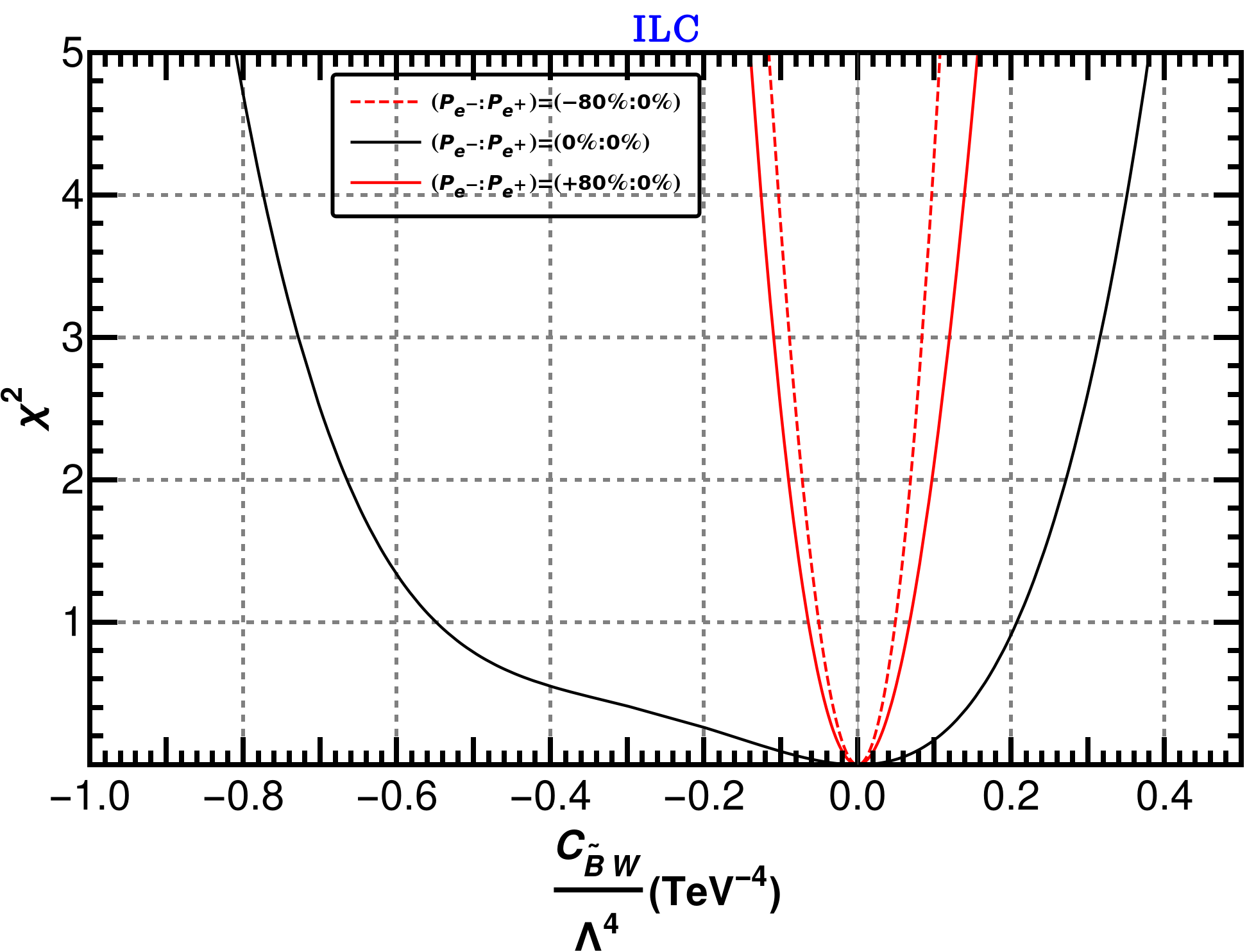}
	\includegraphics[height=6.8cm, width=7.6cm]{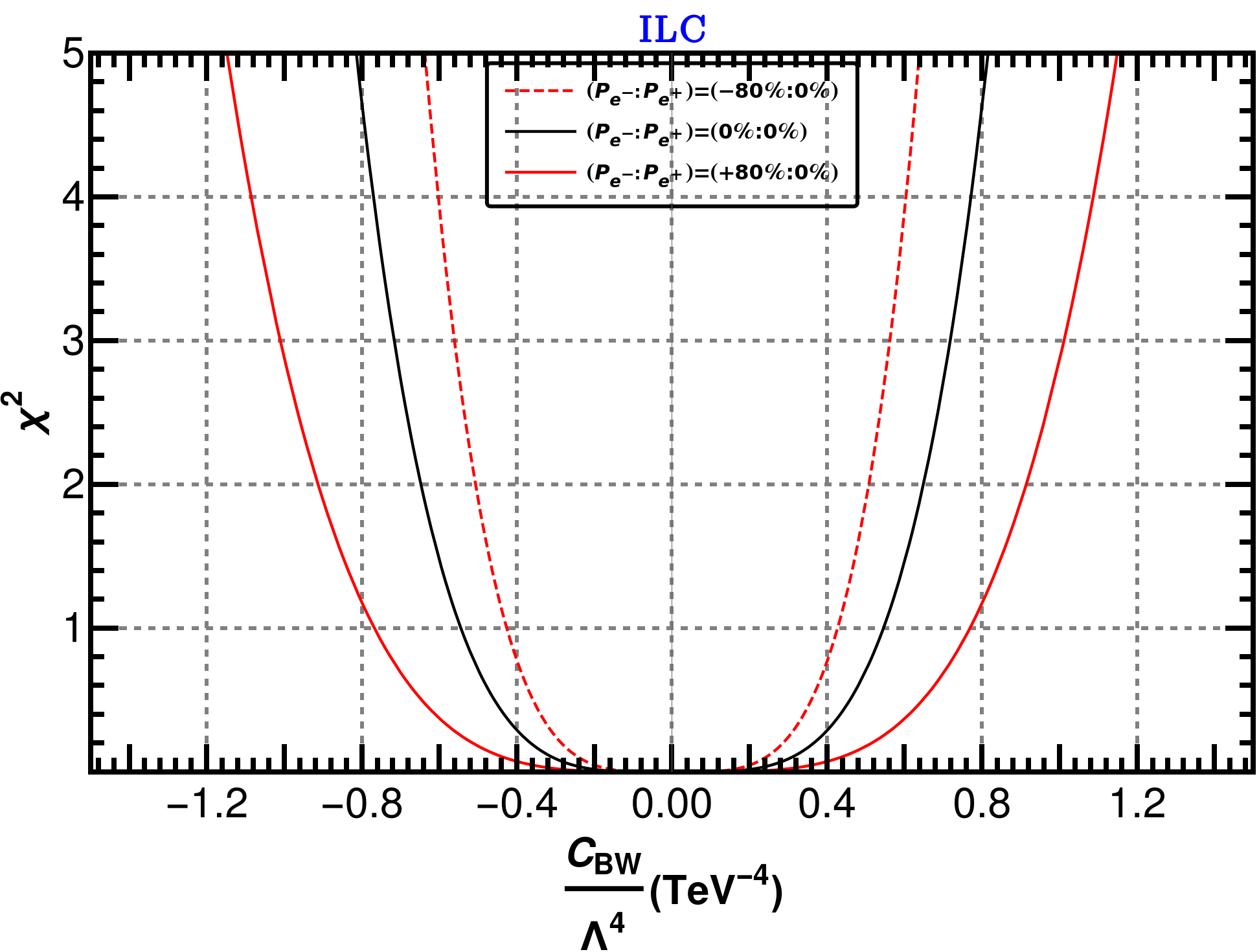}
	$$
	$$
	\includegraphics[height=6.8cm, width=7.6cm]{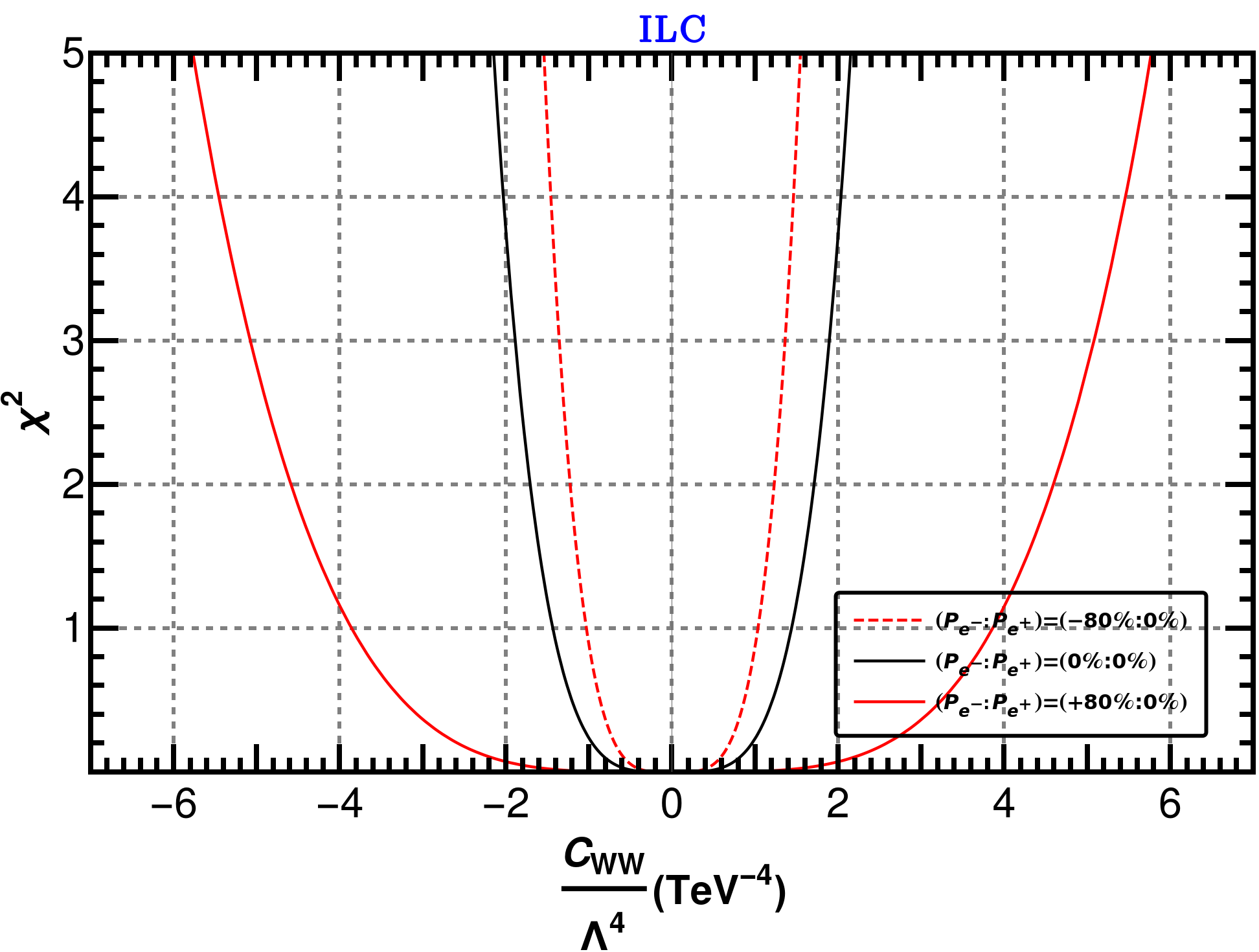}
	\includegraphics[height=6.8cm, width=7.6cm]{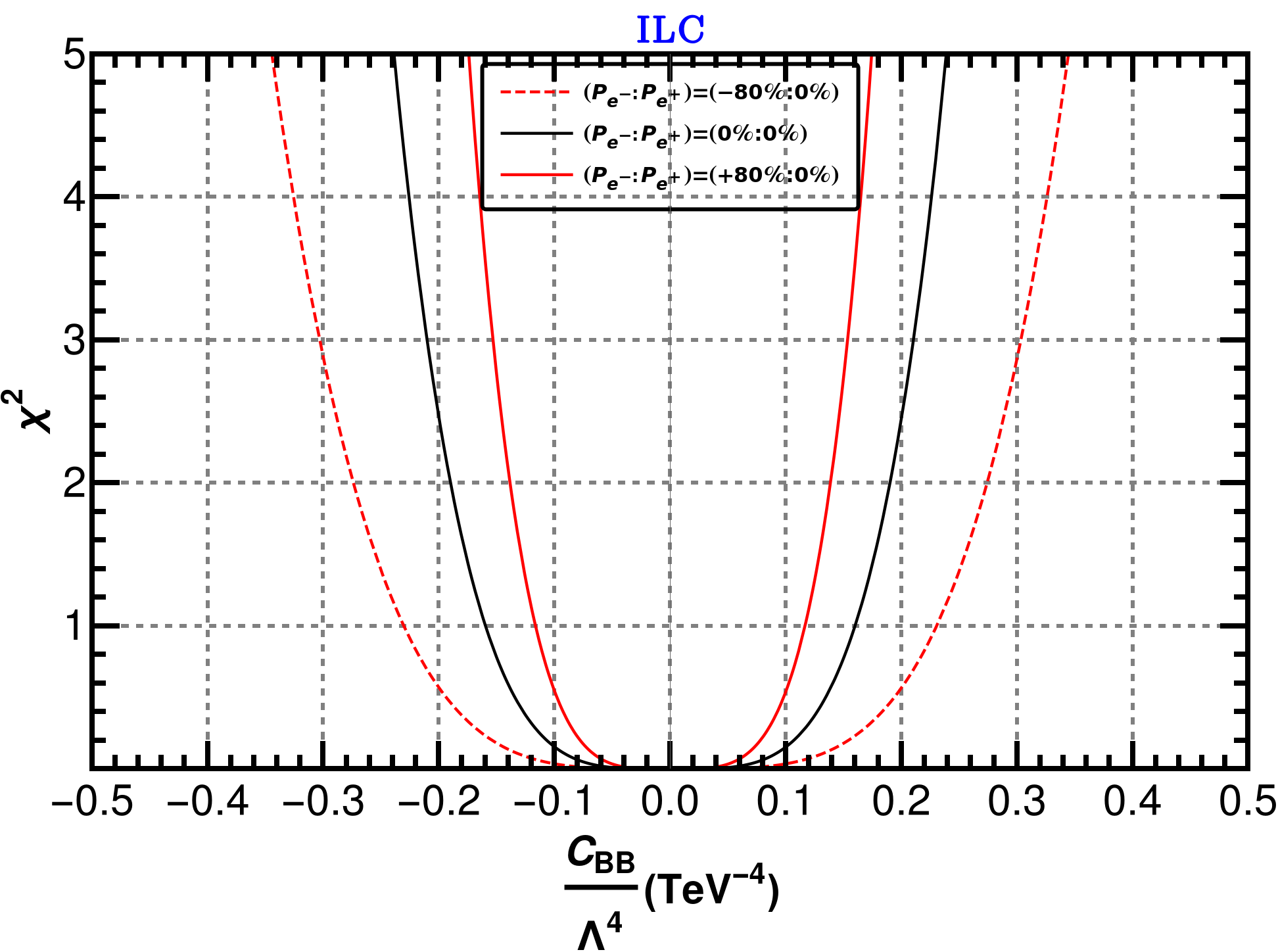}
	$$
	\caption{$\chi^2$ as function of different NP for different choice of polarization combinations in the context of ILC. All the relevant parameters are written in the inset. Top left: $\left(\frac{C_{\tilde B W}}{\Lambda^4}\right)$; top right: $\left(\frac{C_{ B W}}{\Lambda^4}\right)$; bottom left: $\left(\frac{C_{W W}}{\Lambda^4}\right)$; bottom right: $\left(\frac{C_{BB}}{\Lambda^4}\right)$.} 	
	\label{fig:95cl1}
\end{figure}

\begin{figure}[htb!]
	$$
	\includegraphics[height=6.8cm, width=7.6cm]{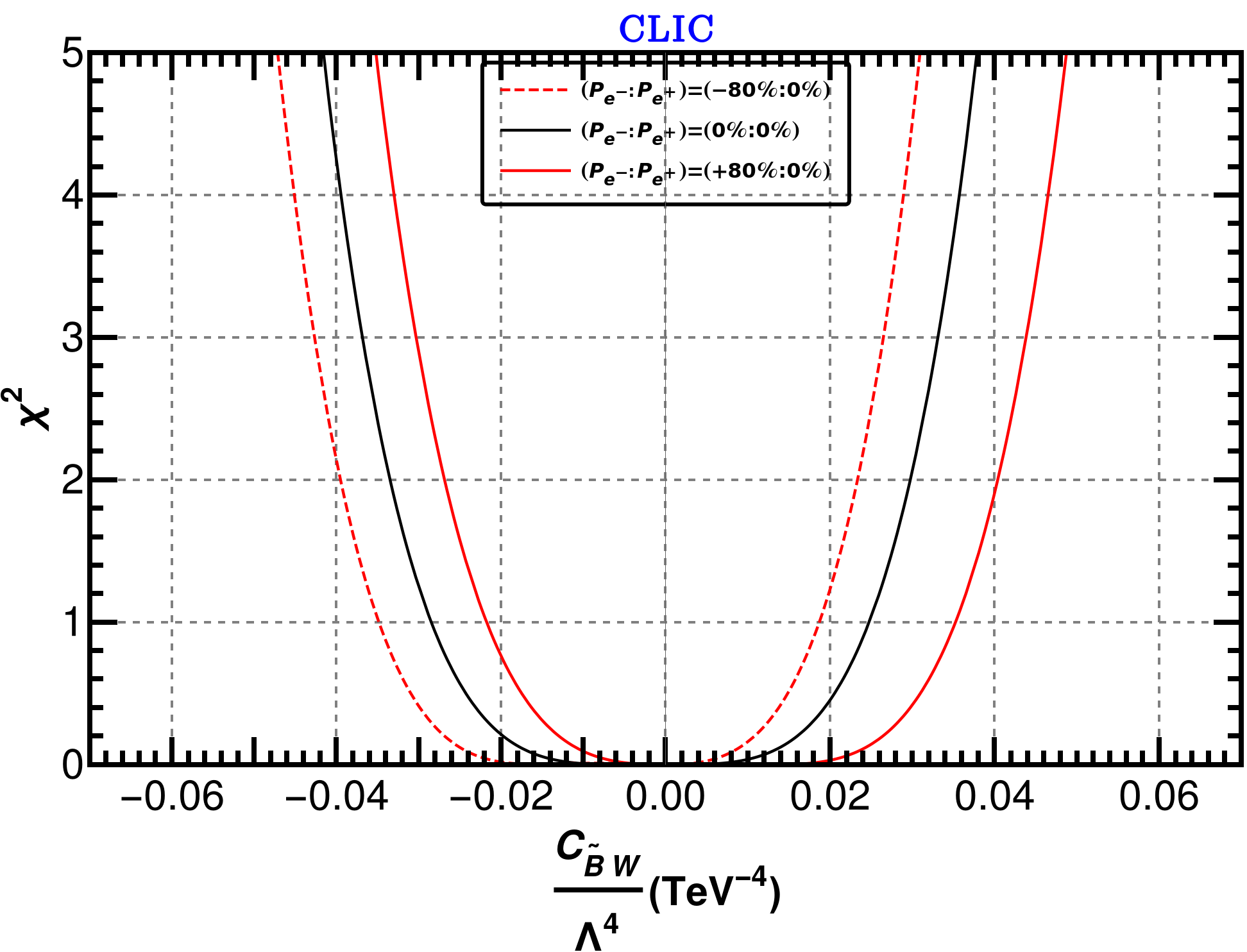}
	\includegraphics[height=6.8cm, width=7.6cm]{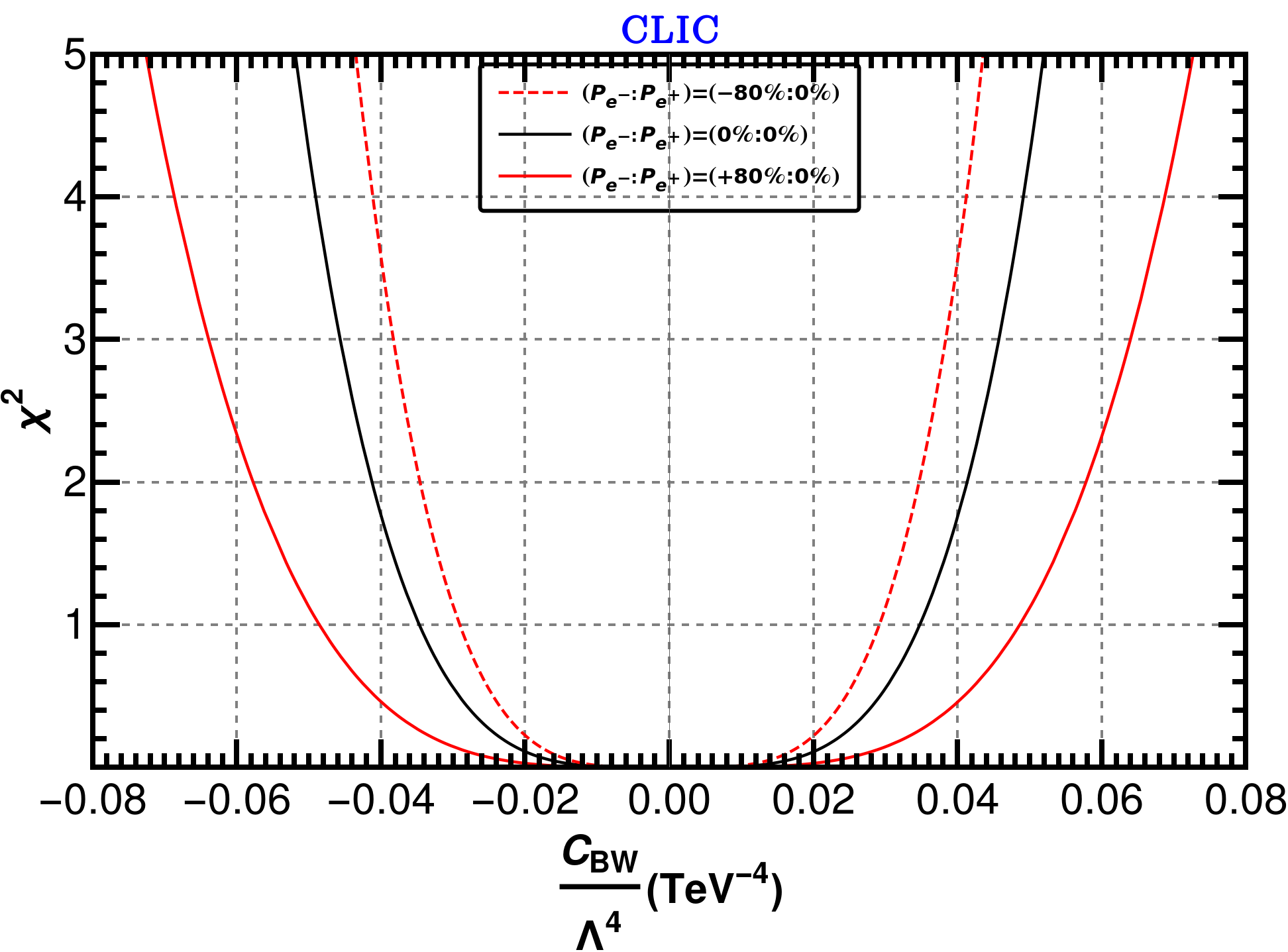}
	$$
	$$
	\includegraphics[height=6.8cm, width=7.6cm]{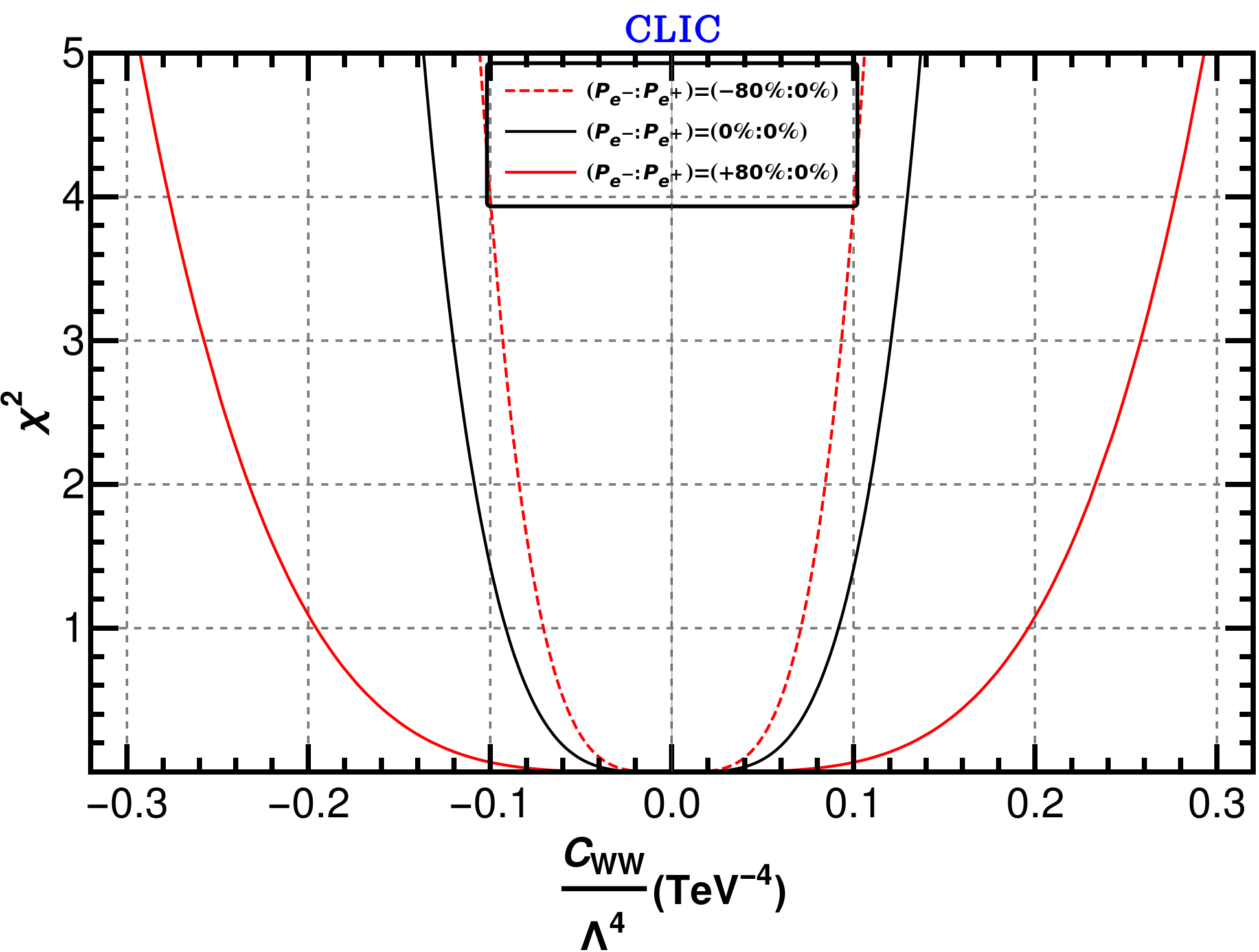}
	\includegraphics[height=6.8cm, width=7.6cm]{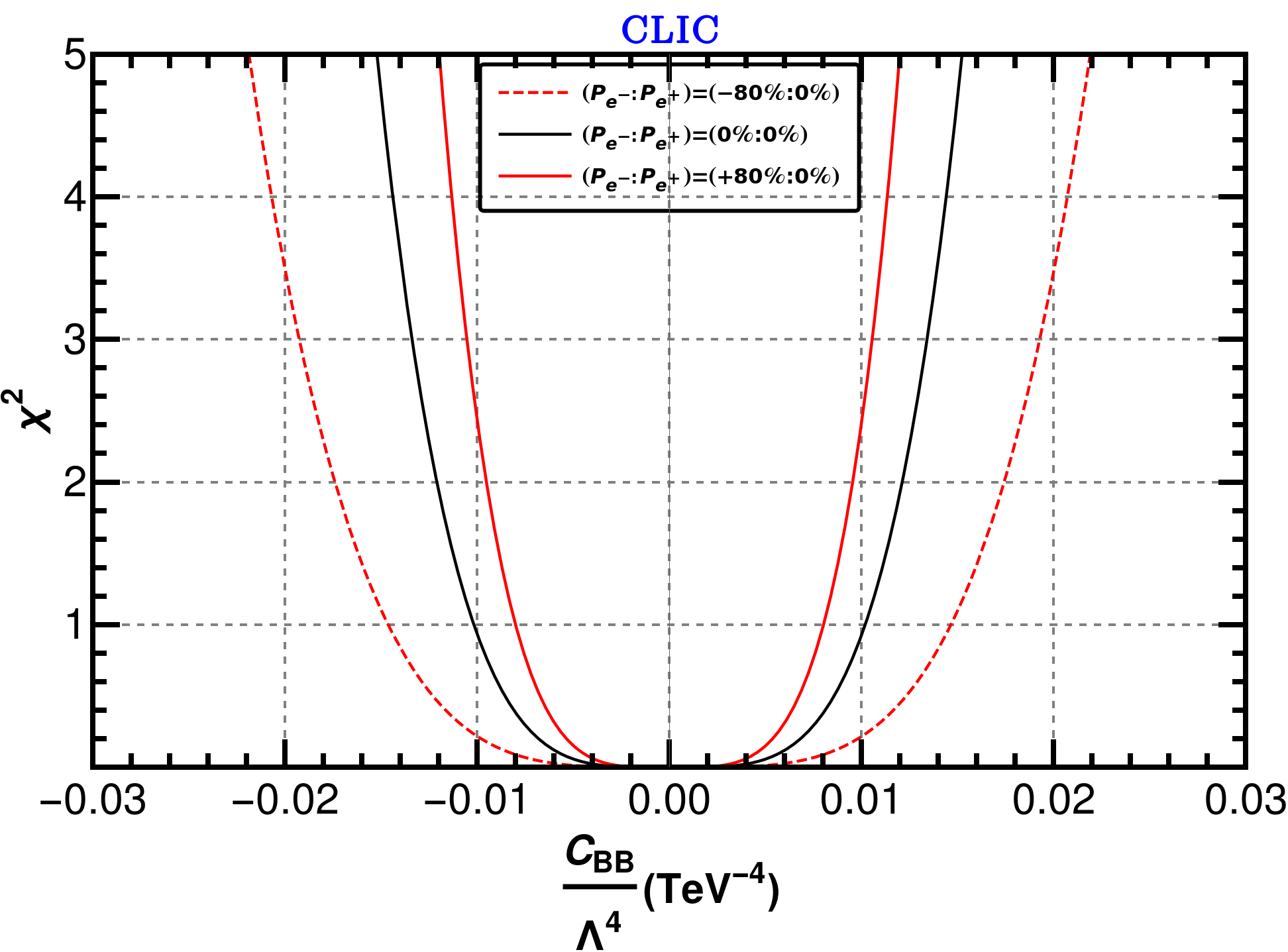}
	$$
	\caption{Same as  figure \ref{fig:95cl1} but for CLIC.} 	
	\label{fig:95cl}
\end{figure}

Using OOT, $95\%$ C.L. limit of NP couplings can be obtained from Eq.~\eqref{eq:chi2}.
We first consider $\sqrt{s}$ = 1 TeV and 3 TeV with $\mathcal{L}_{\tt int}$ = 1000 $\rm fb^{-1}$ and show 1-D $\chi^2$ as function of different NP couplings in the figure \ref{fig:95cl1} and \ref{fig:95cl} respectively.  While evaluating the $\chi^2$ in terms of one NP coupling, the other couplings are kept at zero.  $95\%$ C.L. limit on the NP couplings has been tabulated in Table \ref{tab:95cl}.
We can see that, different NP couplings are sensitive to different choices of polarization combinations, because of unequal contribution of NP couplings to the various helicity amplitudes.
We find that in case of ILC, for $\left(\frac{C_{\tilde B W}}{\Lambda^4}\right)$, $\left(\frac{C_{B W}}{\Lambda^4}\right)$ and $\left(\frac{C_{W W}}{\Lambda^4}\right)$ the best sensitivity is achieved with $\{P_{e^-}:P_{e^+}=-80\%:+30\%\}$, whereas for $\left(\frac{C_{B B}}{\Lambda^4}\right)$, $\{P_{e^-}:P_{e^+}=+80\%:-30\%\}$ produces the best result. In context of CLIC, for  $\left(\frac{C_{B B}}{\Lambda^4}\right)$ we get best sensitivity for $\{P_{e^-}:P_{e^+}=+80\%:0\%\}$ and for $\left(\frac{C_{\tilde B W}}{\Lambda^4}\right)$, $\left(\frac{C_{B W}}{\Lambda^4}\right)$ and $\left(\frac{C_{W W}}{\Lambda^4}\right)$, the best sensitivity is obtained with $\{P_{e^-}:P_{e^+}=-80\%:0\%\}$. This phenomenon can be also understood from figure~\ref{fig:xsecunpol}, where the contribution to $Z\gamma$ production cross-section from $\frac{C_{B B}}{\Lambda^4}$ is maximum in the polarization combination $\{P_{e^-}:P_{e^+}=+80\%:0\%\}$, whereas the other three couplings contribute maximally with $\{P_{e^-}:P_{e^+}=-80\%:0\%\}$.
It is also evident that the initial beam polarization indeed enhances statistical precision in case of all four aNTGCs. 
In Table~\ref{tab:95cl}, we have quoted the statistical uncertainty (95\% C.L) for all the aNTGC couplings for ILC and CLIC and make a comparison between the two. We can see that due to increased signal cross-section, CLIC yields better sensitivity compared to ILC at same integrated luminosity. The comparison between ILC, CLIC and ATLAS sensitivities will be discussed next.

\begin{table}
	\centering
	\begin{tabular}{| c | c |  c | c |  c | c | c | c   c | c  c | c  c| c  c| c c| } 
		\hline
		\multicolumn{1}{|c}{} &
		\multicolumn{6}{|c|}{$95\%$ C.L. limit} \\
		\cline{2-7}
		\multicolumn{1}{|c}{Couplings} &
		\multicolumn{3}{|c}{ILC} & 
		\multicolumn{3}{|c|}{CLIC} \\
		\cline{2-7}
		\multicolumn{1}{|c|}{$\rm (TeV^{-4})$}&
		\multicolumn{1}{c|}{$P_{e^\pm} = ^{00\%}_{00\%}$} &
		\multicolumn{1}{c}{$P_{e^\pm} = ^{ +30\%}_{-80\%}$}&
		\multicolumn{1}{|c|}{$P_{e^\pm} = ^{-30\%}_{+80\%}$}&
		\multicolumn{1}{c}{$P_{e^\pm} = ^{00\%}_{00\%}$}&
		\multicolumn{1}{|c}{$P_{e^\pm} = ^{+00\%}_{-80\%}$}&
		\multicolumn{1}{|c|}{$P_{e^\pm} = ^{+00\%}_{+80\%}$}\\
		\hline
		\multirow{2}*{$\frac{C_{\tilde{B}W}}{\Lambda^4}$}& $+0.351$ & $+0.096$ & $+0.139$ & $+0.036$ & $+0.028$  & $+0.046$  \\
		& $-0.774$ & $-0.103$ & $-0.126$ & $-0.039$ & $-0.045$ & $-0.033$ \\
		\hline
		\multirow{2}*{$\frac{C_{BW}}{\Lambda^4}$} & $+0.770$  & $+0.603$  & $+1.086$  & $+0.049$  & $+0.041$ & $+0.069$ \\
		&$-0.770$ & $-0.603$ & $-1.086$ & $-0.049$ & $-0.041$ & $-0.069$ \\
		\hline
		\multirow{2}*{$\frac{C_{WW}}{\Lambda^4}$}&  $+2.032$ & $+1.460$ & $+5.457$ &  $+0.129$ &  $+0.100$ &  $+0.277$ \\
		&$-2.032$ & $-1.460$ &  $-5.457$ & $-0.129$ & $-0.100$ & $-0.277$ \\
		\hline
		\multirow{2}*{$\frac{C_{BB}}{\Lambda^4}$}&  $+0.226$ & $+0.326$ &  $+0.165$ &  $+0.014$ &  $+0.021$  & $+0.011$ \\
		& $-0.226$ & $-0.326$ & $-0.165$ & $-0.014$ & $-0.021$ & $-0.011$ \\
		\hline
	\end{tabular}
	\caption{Optimal 95\% C.L. limit on dimension-8 aNTGCs at ILC and CLIC for different beam polarization combinations.}
	\label{tab:95cl}
\end{table}

\begin{figure}[htb!]
	\begin{align*}
		\includegraphics[height=5cm, width=5cm]{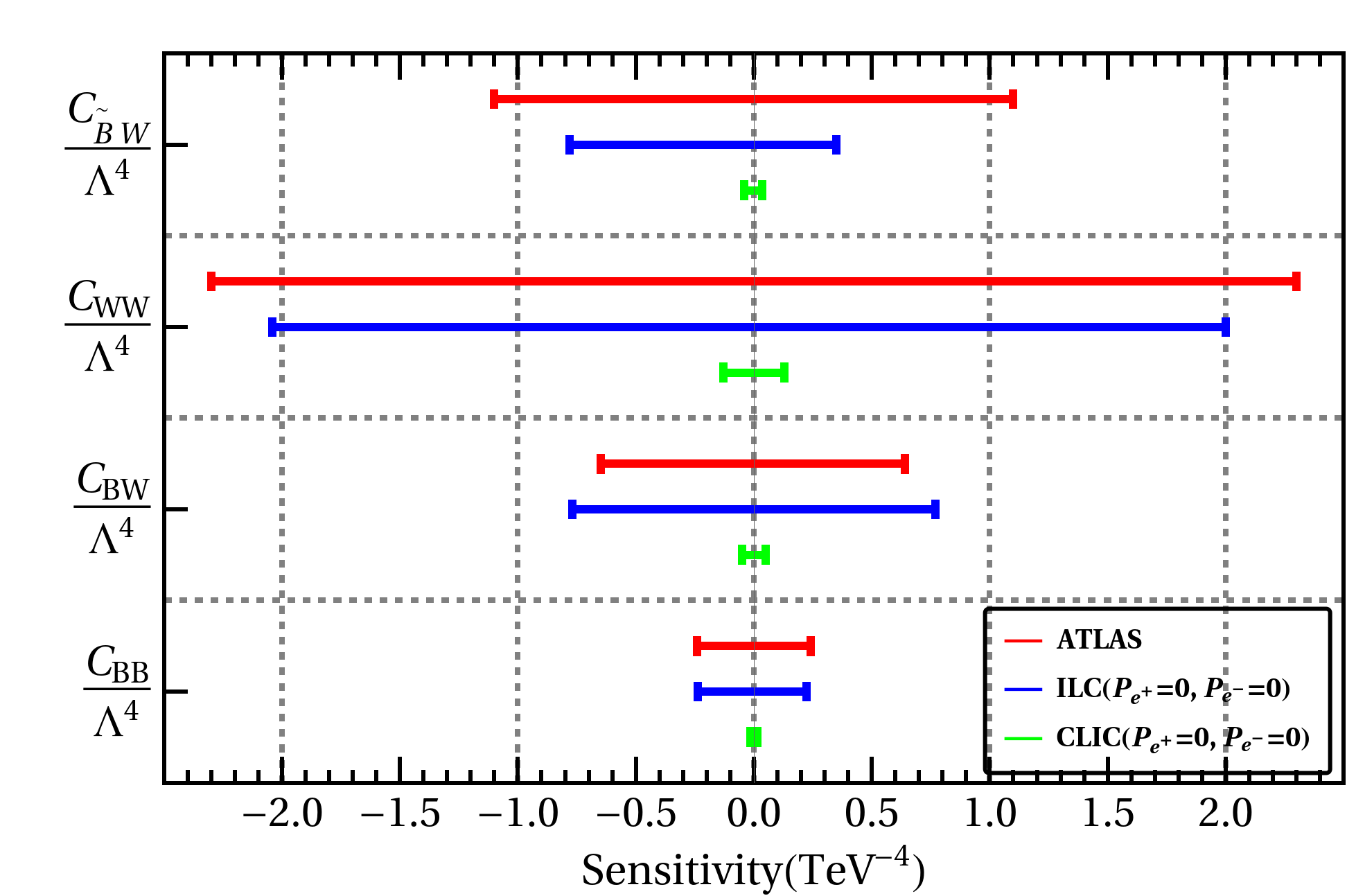} \quad
		\includegraphics[height=5cm, width=5cm]{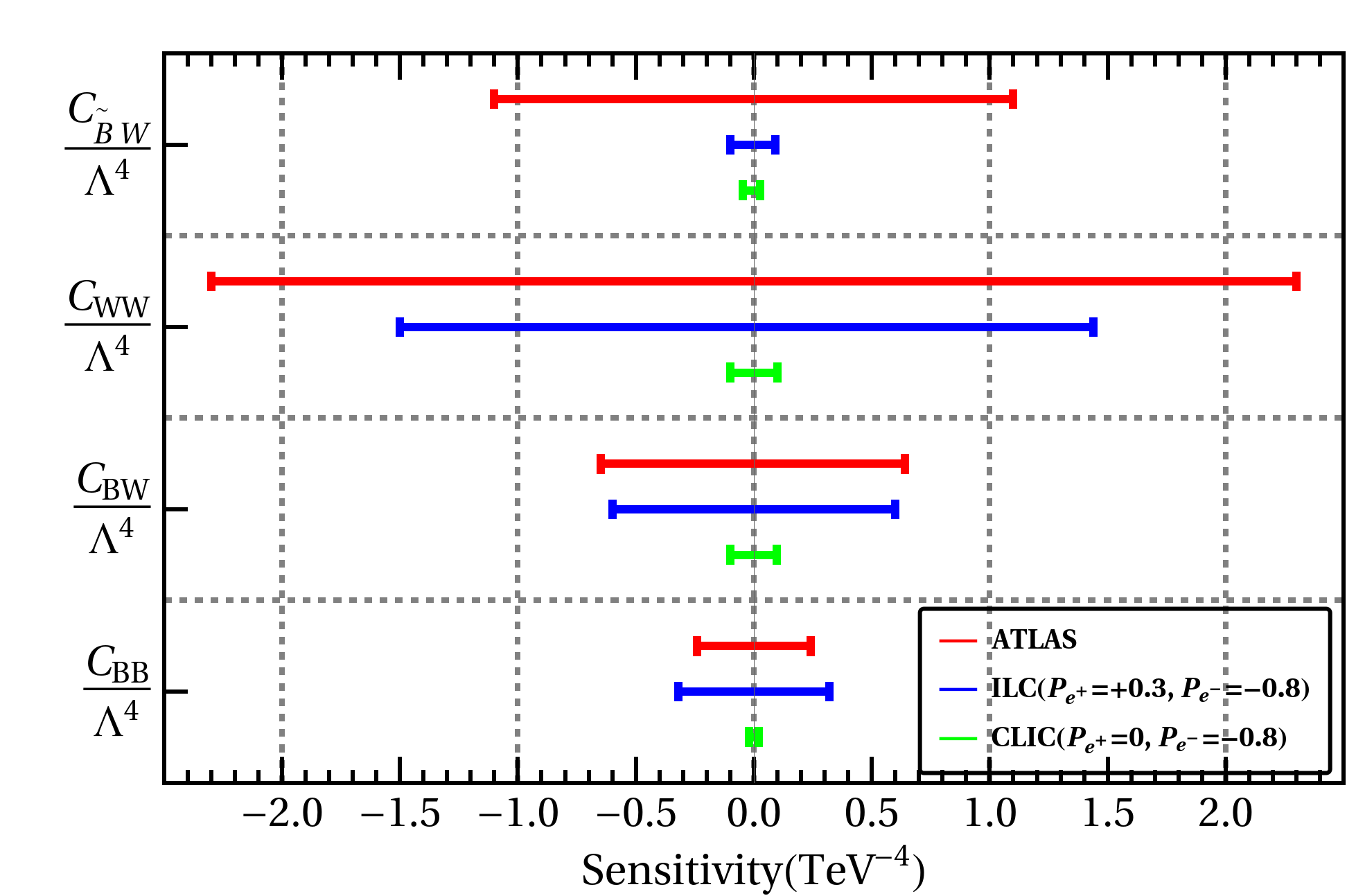} \quad
		\includegraphics[height=5cm, width=5cm]{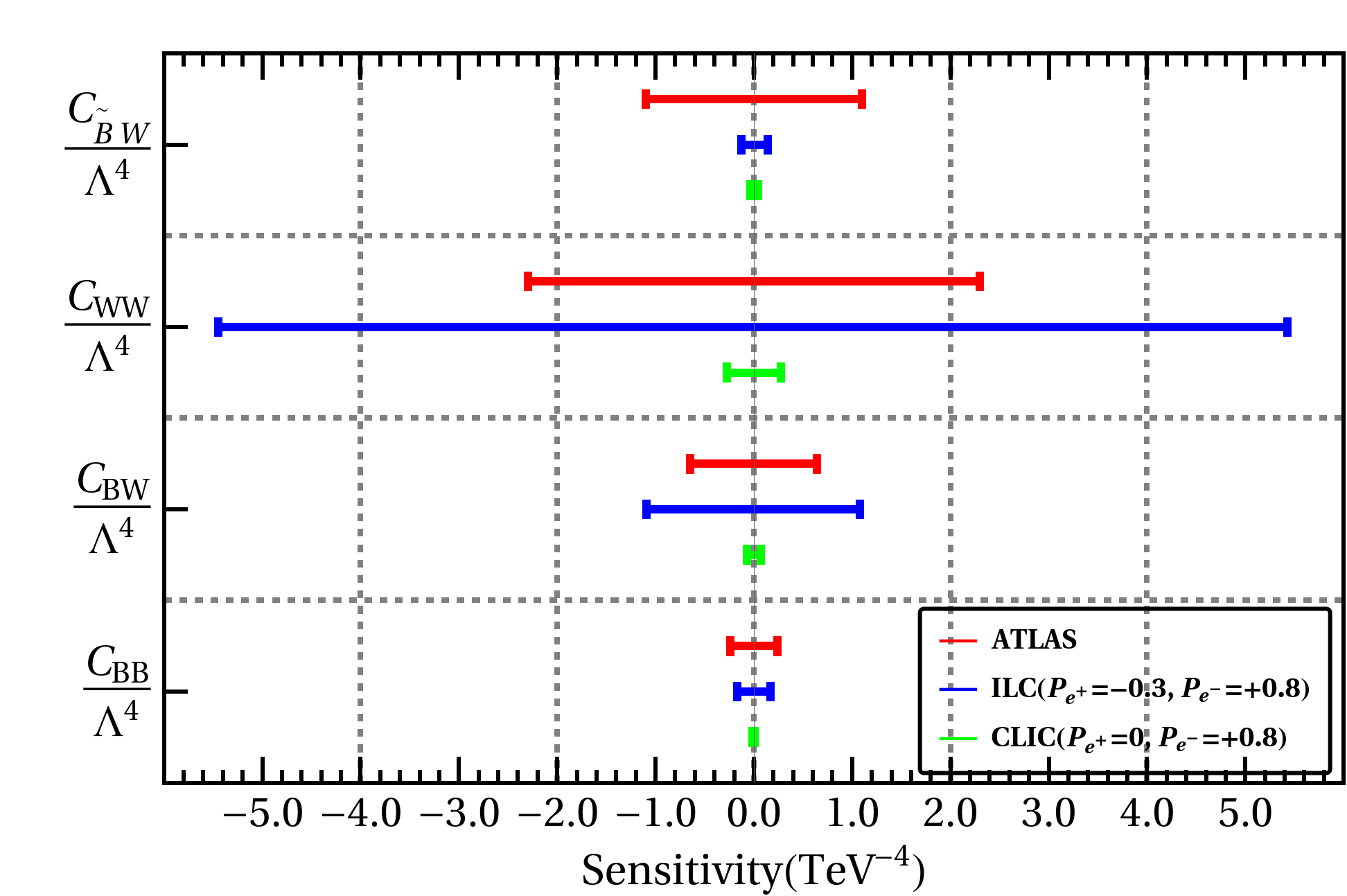} \quad
	\end{align*}
	\caption{Comparison of 95\% C.L. limit of aNTGCs for three different polarization combinations, in the context of ILC and CLIC with the existing ATLAS bound. Left: Unpolarized beams; Middle: electron beams are left polarized; Right: electron beams are right polarized. Detailed polarization information are given in the inset.} 	
	\label{fig:senscomp}
\end{figure}

In figure~\ref{fig:senscomp}, we compare the sensitivities achievable at ILC and CLIC with various polarizations and compare them with the limits obtained with the most recent ATLAS data~\cite{ATLAS:2018nci}. One can see that CLIC outperforms both LHC and ILC by at least an order of magnitude for all polarization combinations. However, in case of ILC, for $\big(\frac{C_{BW}}{\Lambda^4}\big)$, $\big(\frac{C_{WW}}{\Lambda^4}\big)$ and $\big(\frac{C_{\tilde B W}}{\Lambda^4}\big)$ polarization combination $\{P_{e^-}:P_{e^+}=-80\%:+30\%\}$ produce better sensitivity compared to ATLAS, whereas for $\big(\frac{C_{BB}}{\Lambda^4}\big)$, ILC sensitivities are better than that from ATLAS  with $\{P_{e^-}:P_{e^+}=+80\%:-30\%\}$. It is evident from the results that initial beam polarization can improve the level of precision in aNTGCs measurement to a large extent.

\begin{figure}[htb!]
	$$
	\includegraphics[height=7cm, width=7.7cm]{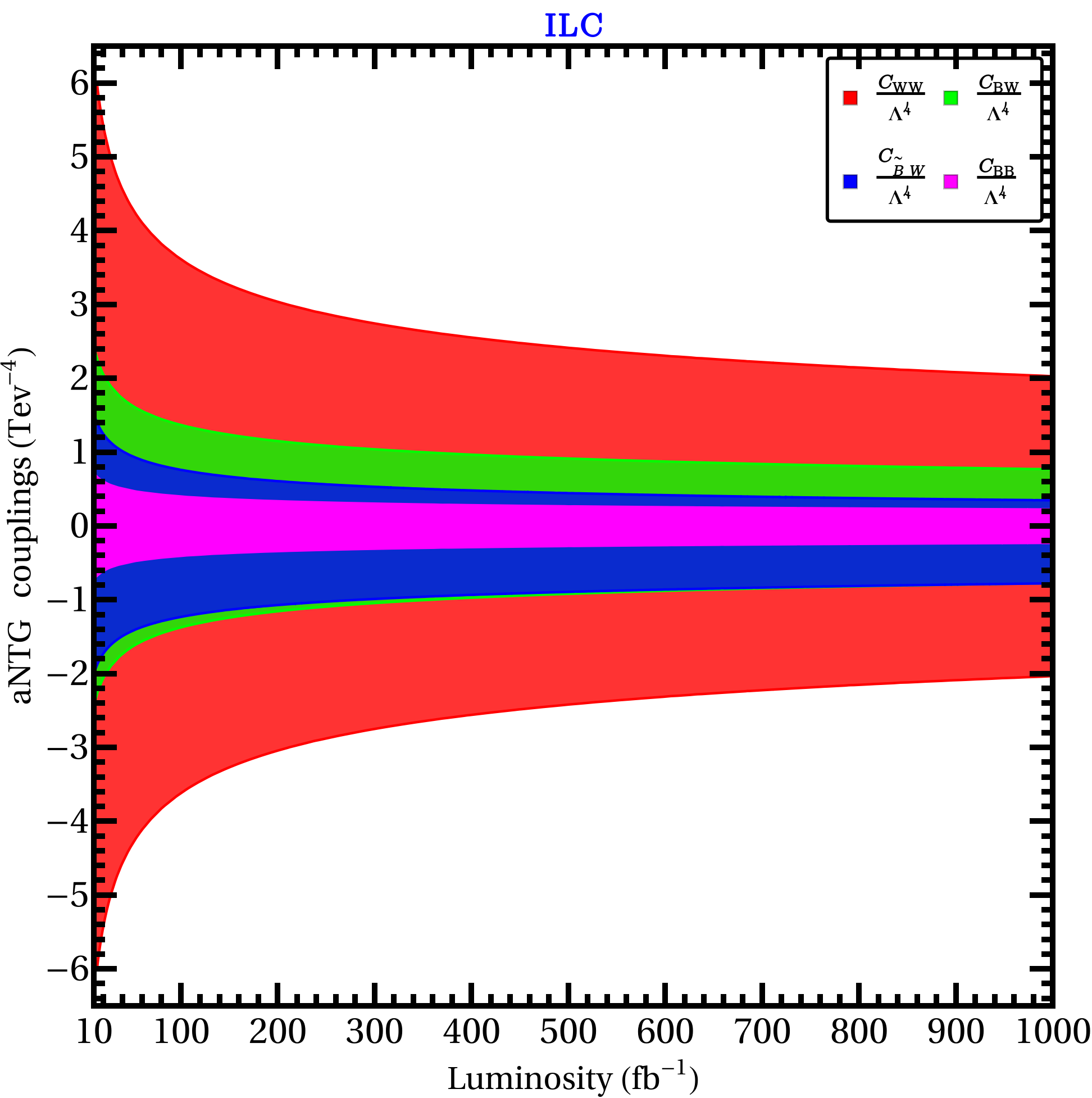}
	\includegraphics[height=7cm, width=7.7cm]{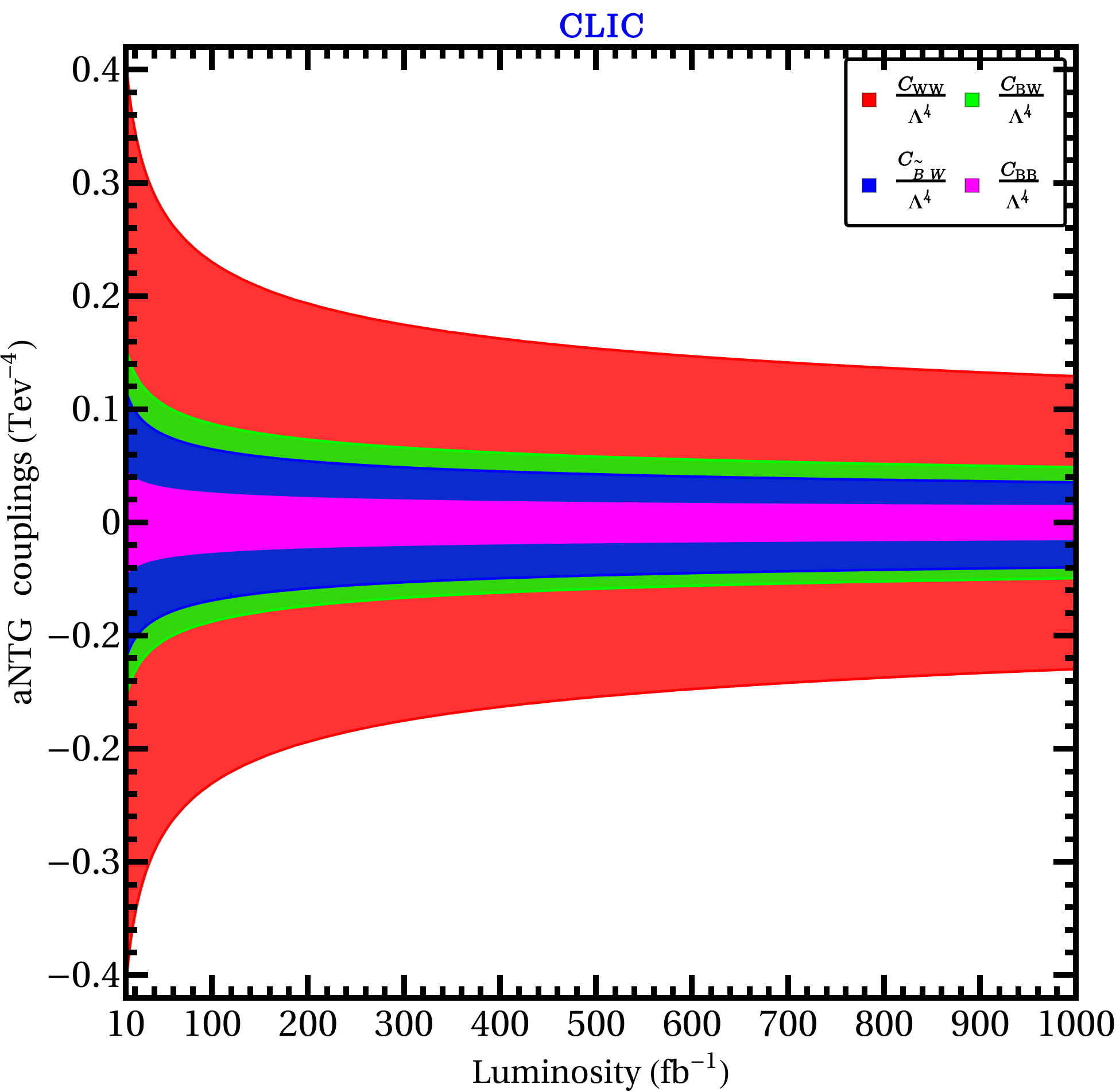}
	$$
	\caption{Variation of the limits on NP parameters with the integrated luminosity at $e^+e^-$ colliders. Left: ILC; Right: CLIC.} 	
	\label{fig:lumvar}
\end{figure} 

\begin{figure}[htb!]
	$$
	\includegraphics[height=4.5cm, width=5.1cm]{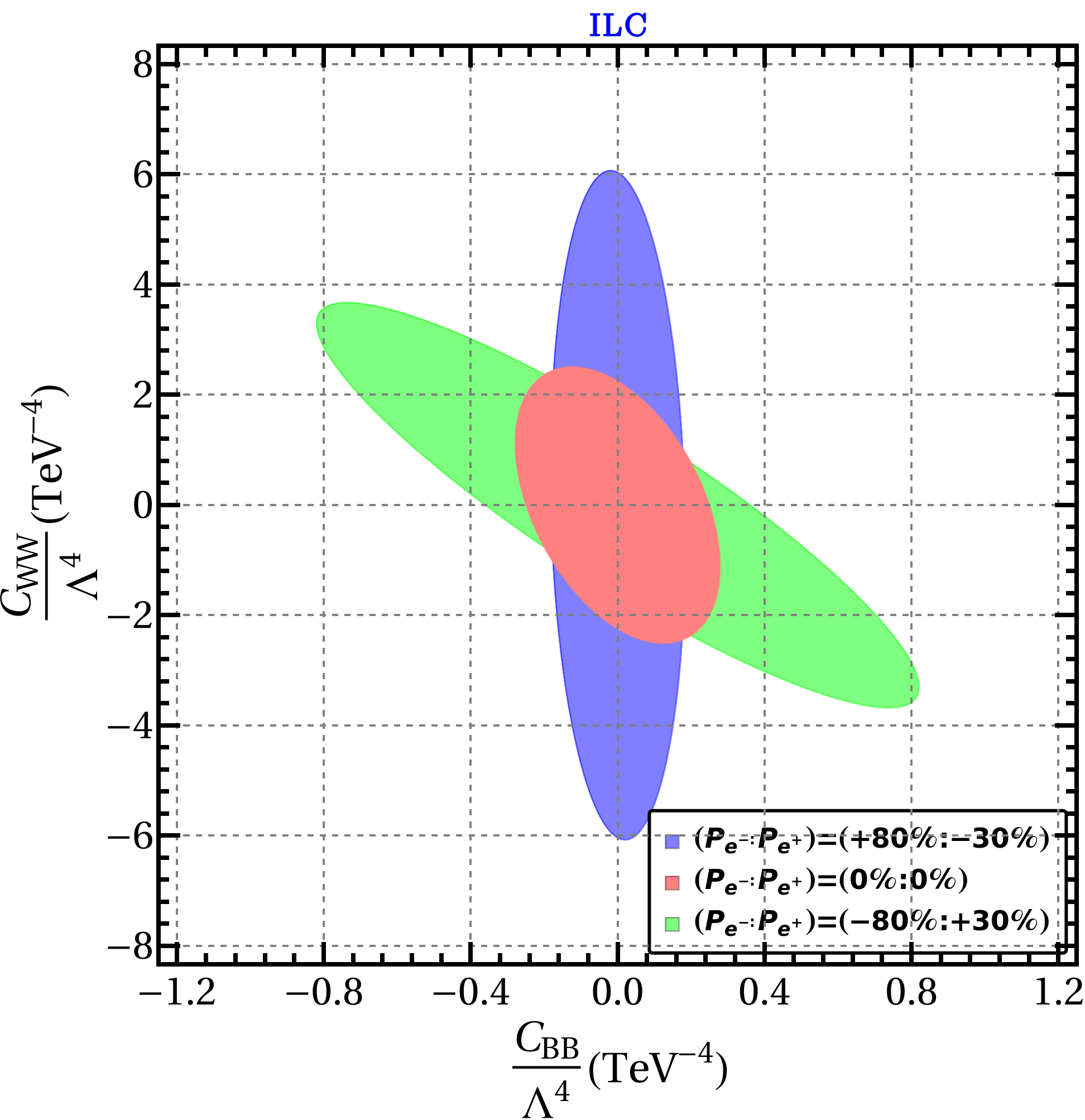}
	\includegraphics[height=4.5cm, width=5.1cm]{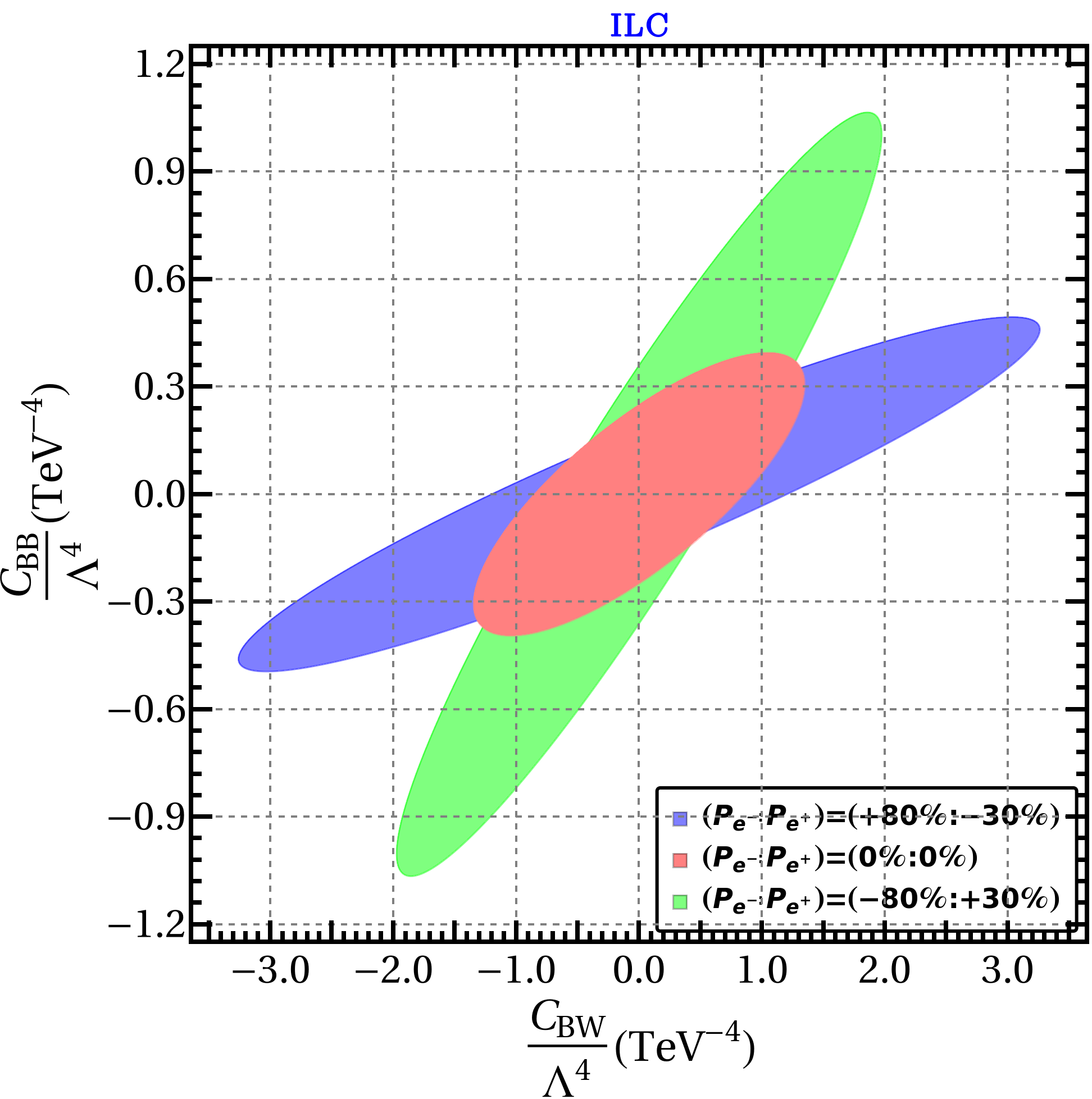}
	\includegraphics[height=4.5cm, width=5.1cm]{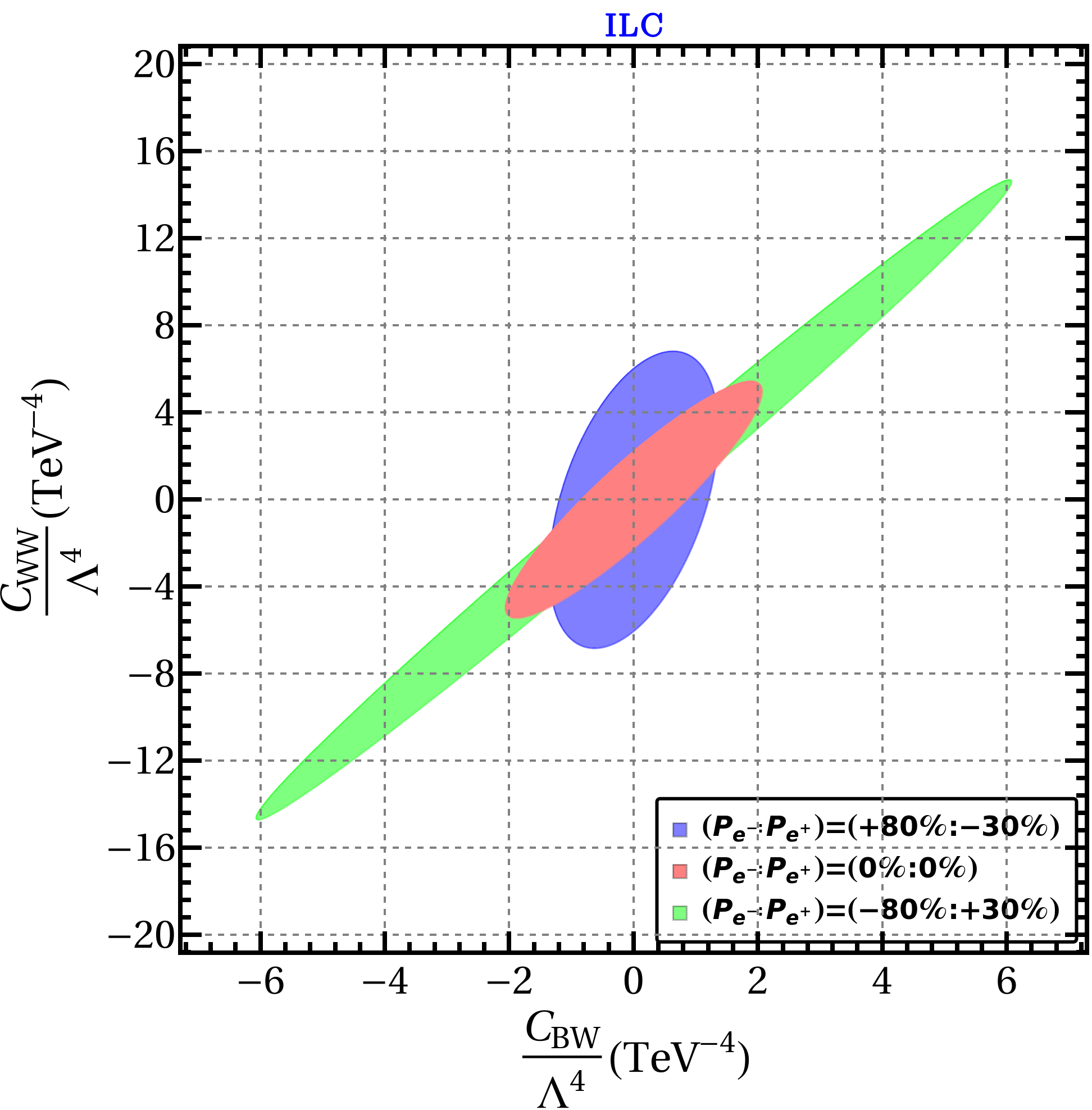}
	$$
	$$
	\includegraphics[height=4.5cm, width=5.1cm]{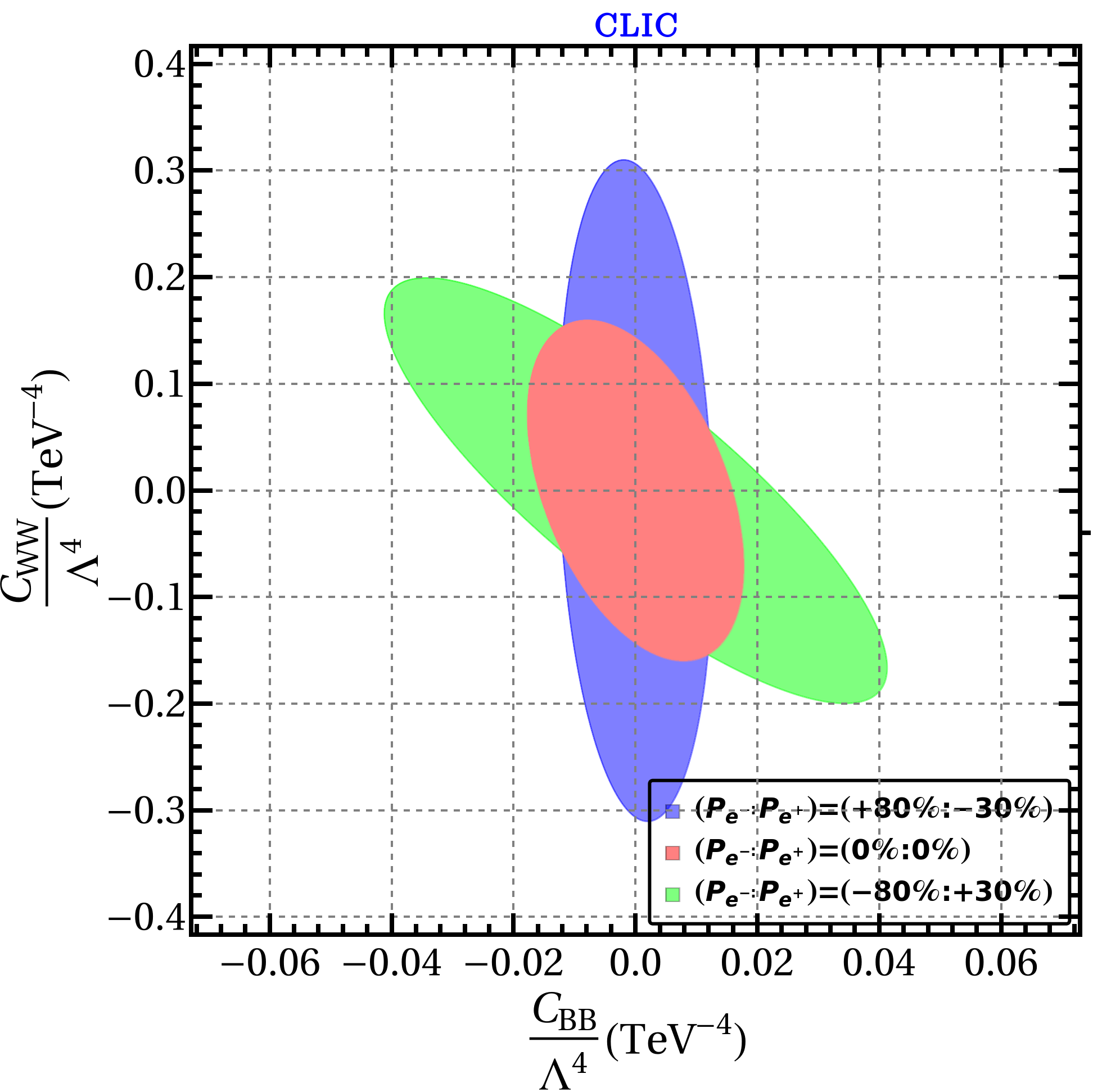}
	\includegraphics[height=4.5cm, width=5.1cm]{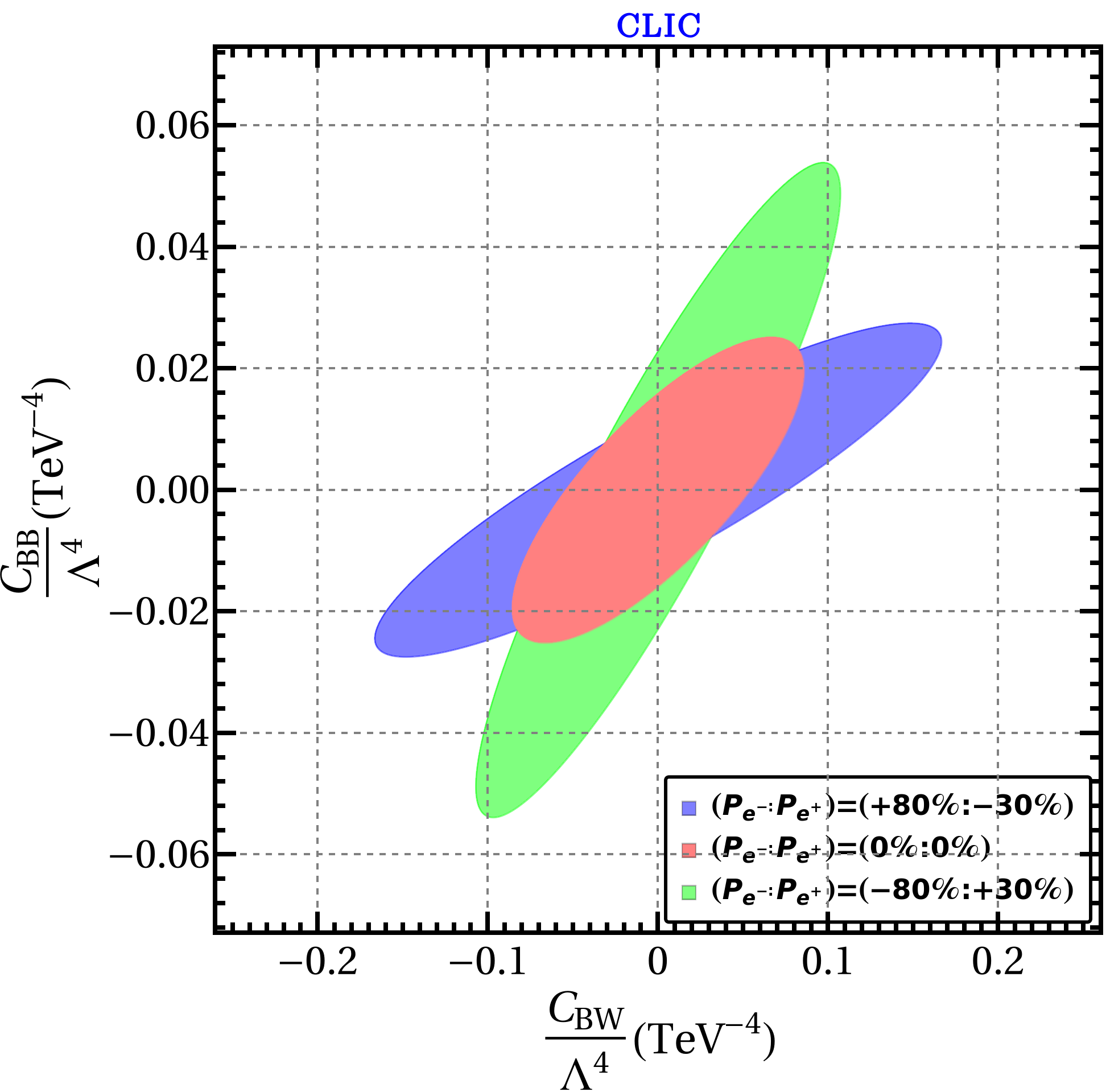}
	\includegraphics[height=4.5cm, width=5.1cm]{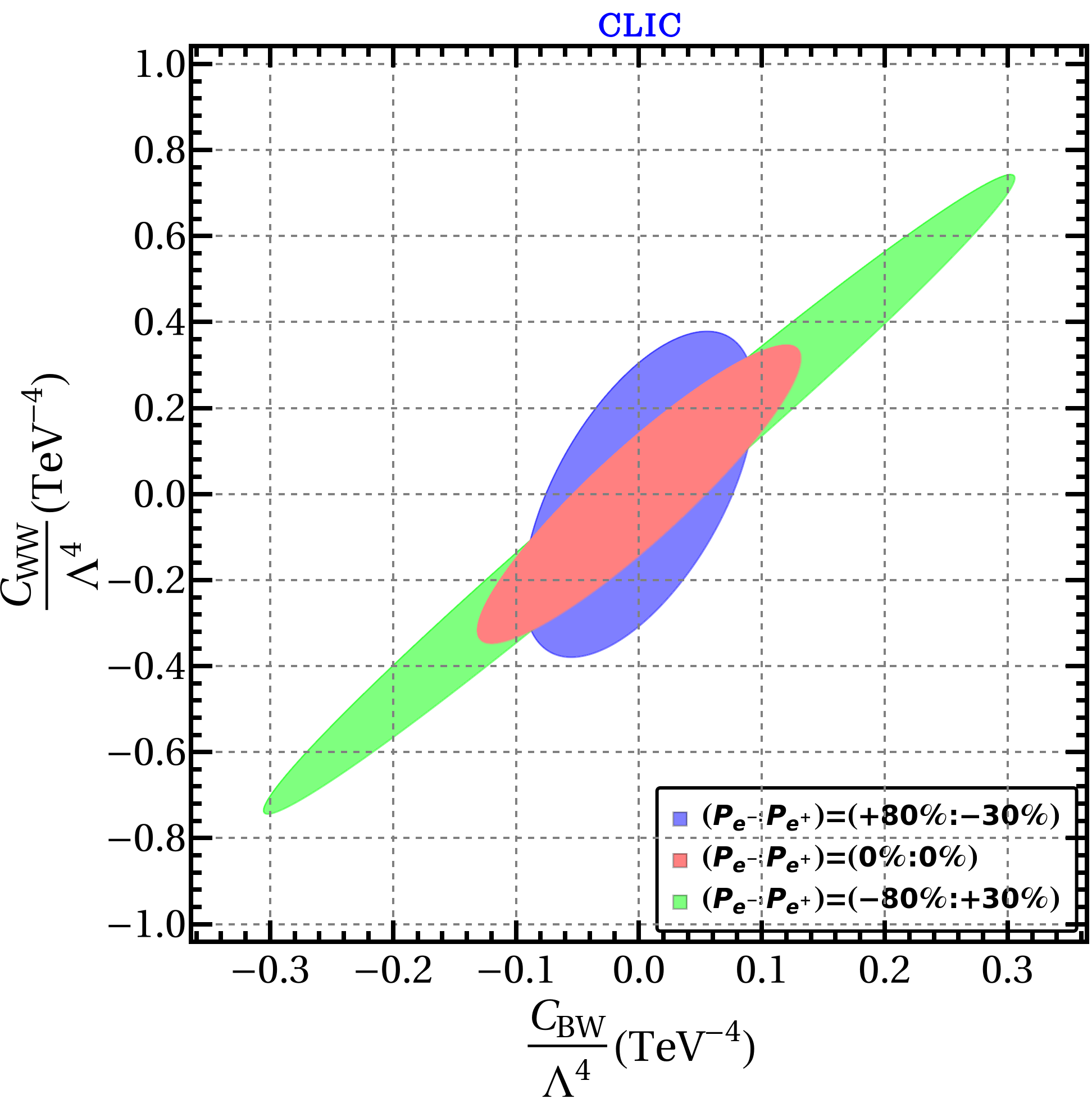}
	$$
	\caption{95\% C.L. limits of CP-violating anomalous couplings in 2D parameter space. Left panel: $\left(\frac{C_{W W}}{\Lambda^4}\right)$ \& $\left(\frac{C_{BB}}{\Lambda^4}\right)$; middle panel: $\left(\frac{C_{BW}}{\Lambda^4}\right)$ \& $\left(\frac{C_{BB}}{\Lambda^4}\right)$.\; right panel:$\left(\frac{C_{BW}}{\Lambda^4}\right)$ \& $\left(\frac{C_{WW}}{\Lambda^4}\right)$. Top:ILC; Bottom:CLIC.} 	
	\label{fig:95cl2d}
\end{figure}

In figure~\ref{fig:lumvar}, we show the variation of statistical limits (95\% C.L.) of the NP parameters with the integrated luminosity in the context of both ILC and CLIC. We have allowed the luminosity to vary from 10 $\rm fb^{-1}$ to 1000 $\rm fb^{-1}$. With the increasing luminosity, the statistical limit of all the NP parameters is decreased and follow a similar trend. Varying the luminosity from 10 $\rm fb^{-1}$ to 1000 $\rm fb^{-1}$ enhances the statistical limits by 76\% (64\%), 69\% (67\%), 67\% (65\%), and 67\% (64\%) for $\left(\frac{C_{\tilde{B}W}}{\Lambda^4}\right)$, $\left(\frac{C_{BW}}{\Lambda^4}\right)$, $\left(\frac{C_{WW}}{\Lambda^4}\right)$, and $\left(\frac{C_{BB}}{\Lambda^4}\right)$, respectively for ILC (CLIC). 

\subsection{Correlation of CP-violating aNTGCs}
\label{sec:corel}

Having discussed our results for each individual NP couplings, we will present our two-parameter analysis. The bounds shown in Figure~\ref{fig:senscomp} are the most conservative ones. However, there is a possibility that more than one non-zero couplings participate in $e^+e^- \rightarrow Z\gamma$ process. For example, when CP-violation is considered, $\big(\frac{C_{B W}}{\Lambda^4}\big)$, $\big(\frac{C_{W W}}{\Lambda^4}\big)$ and $\big(\frac{C_{B B}}{\Lambda^4}\big)$ can contribute simultaneously in $Z\gamma$ production and therefore, the sensitivity in measuring one coupling can be affected by the measurement of the other coupling. In order to capture this non-trivial correlation between these couplings, we have considered two aNTGCs non-zero at a time and computed the bound on the parameter space spanned by them. In figure~\ref{fig:95cl2d}, we show the statistical limits on two-parameters space spanned by $\{\frac{C_{BB}}{\Lambda^4}, \frac{C_{WW}}{\Lambda^4}\}$, $\{\frac{C_{BB}}{\Lambda^4}, \frac{C_{BW}}{\Lambda^4}\}$ and $\{\frac{C_{WW}}{\Lambda^4}, \frac{C_{BW}}{\Lambda^4}\}$ respectively for different polarization combinations. Like one-parameter analysis, here too we can see that CLIC can offer precision at least one order of magnitude higher than that of ILC. Interestingly, we see non-trivial correlation between two parameters, which vary with varying degree of polarization. This happens due to non-trivial interference terms between various couplings. We see, unlike the one-parameter case, here even unpolarized beams can give rise to best sensitivity along certain coupling directions. Therefore, one can say, the two-parameter analysis can provide significantly different sensitivity and polarization-dependence, again owing its origin to the constructive or destructive interference.

In the context of ILC, we have considered the position beam to be both right and left polarized with same degree ($P_{e^+}=\pm 30\%$) whereas for CLIC the positron beam is unpolarized, while in both cases electron beam is supposed to have a higher degree of polarization ($P_{e^-}=\pm 80\%$). Now, from figure \ref{fig:95cl2d}, we can see that the is no significant change in the nature of correlation of aNTGCs in two-parameters space if we compare the elliptic contours between ILC and CLIC. However, the actual sizes of the contours will definitely depend on the degree of polarization. Therefore, one can infer that the correlation between two couplings will be dominantly governed by on electron beam polarization in case of both ILC and CLIC at the design polarization.

\subsection{Sensitivity comparison: OOT vs cut-based analysis}
\label{sec:ootvscol}
\begin{figure}[htb!]
	$$
	\includegraphics[height=4.5cm, width=5.1cm]{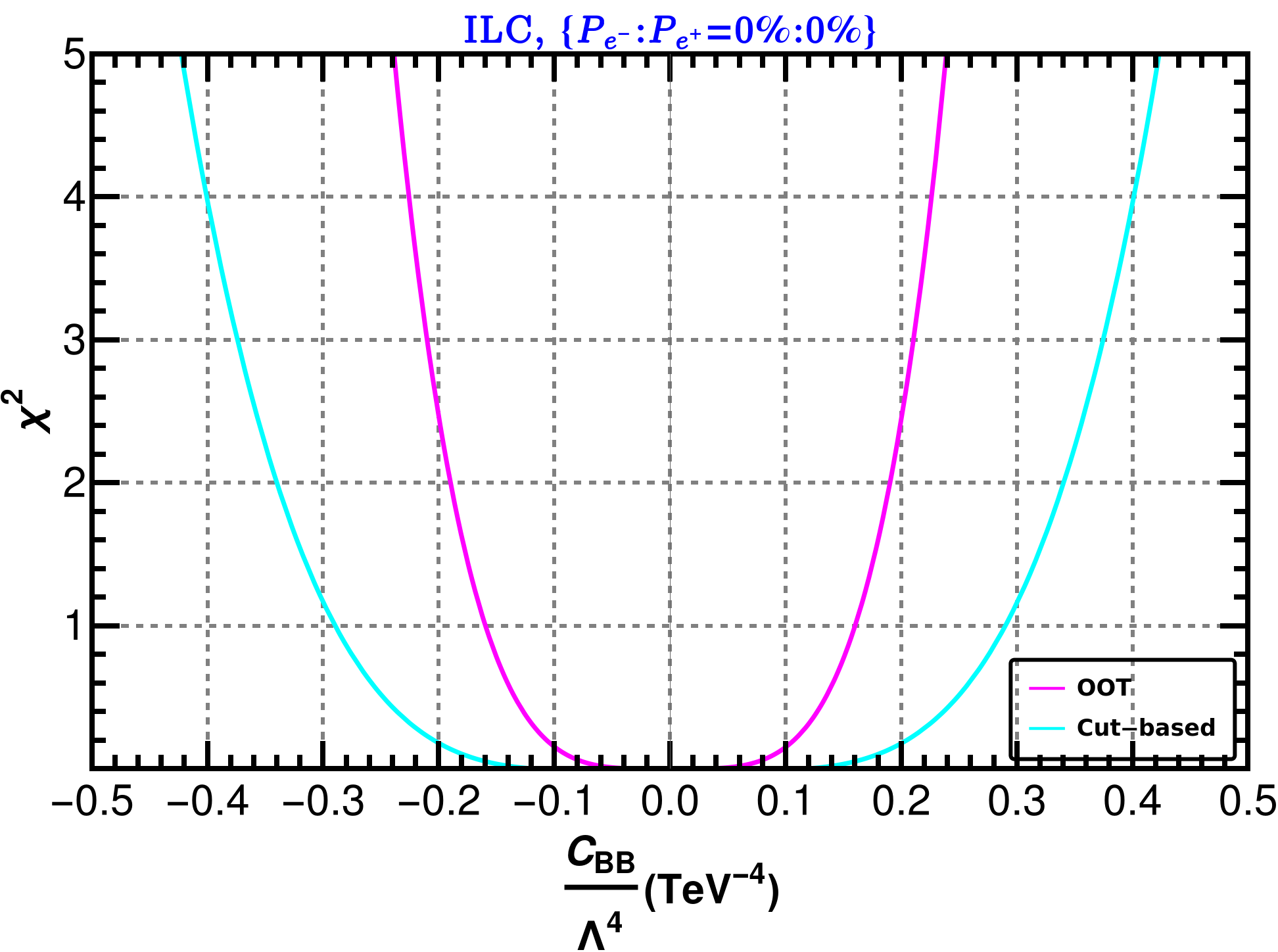}
	\includegraphics[height=4.5cm, width=5.1cm]{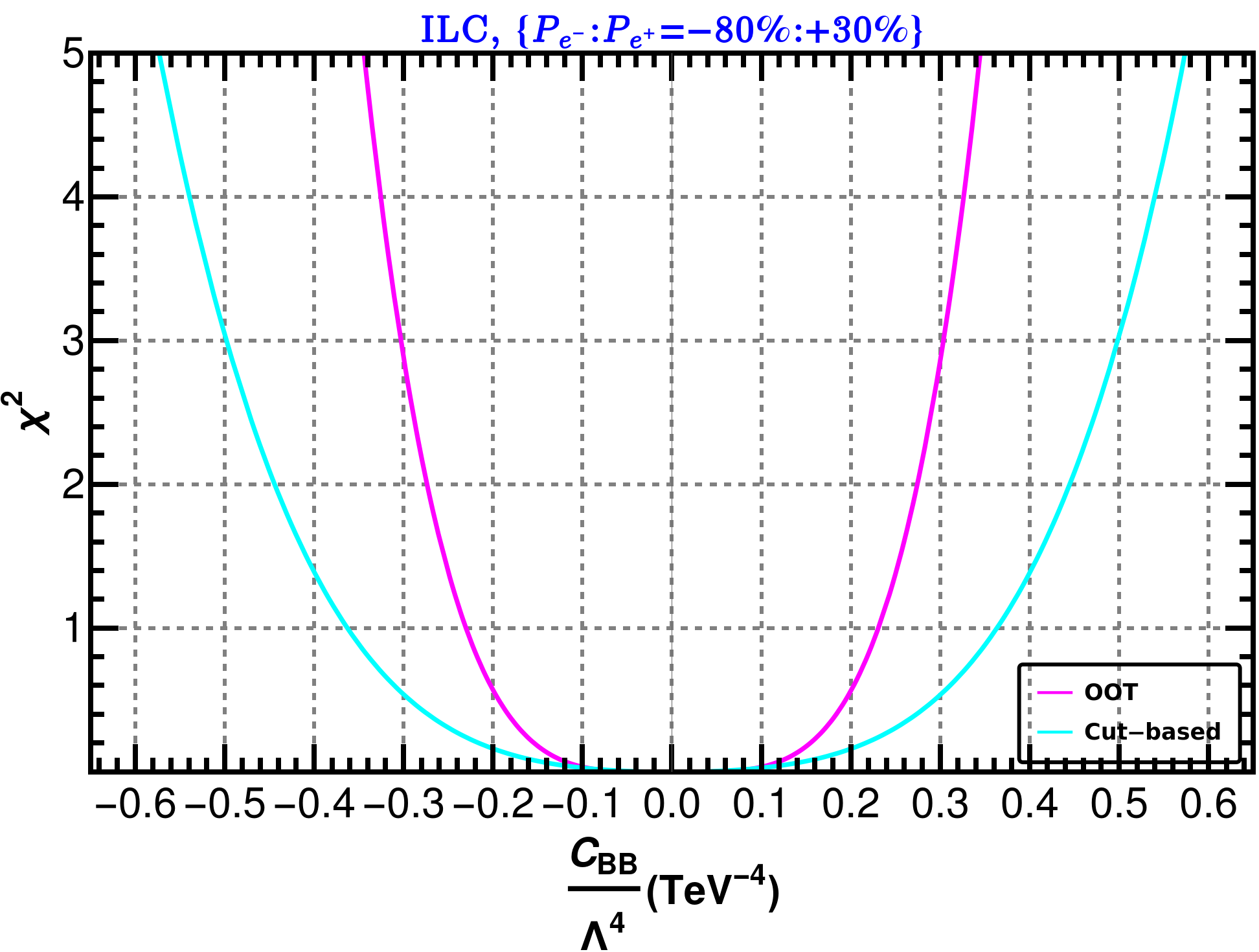}
	\includegraphics[height=4.5cm, width=5.1cm]{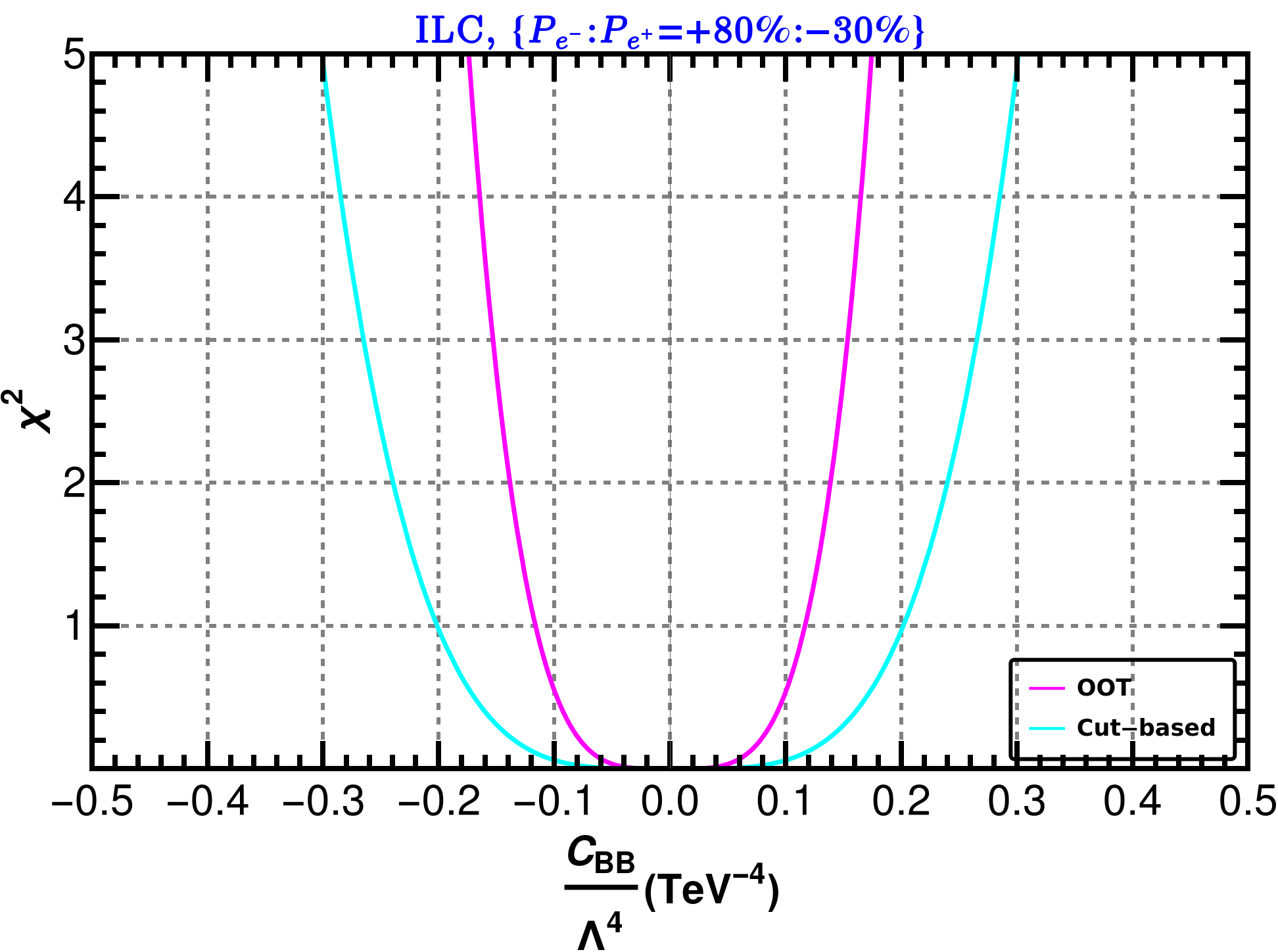}
	$$
	$$
	\includegraphics[height=4.5cm, width=5.1cm]{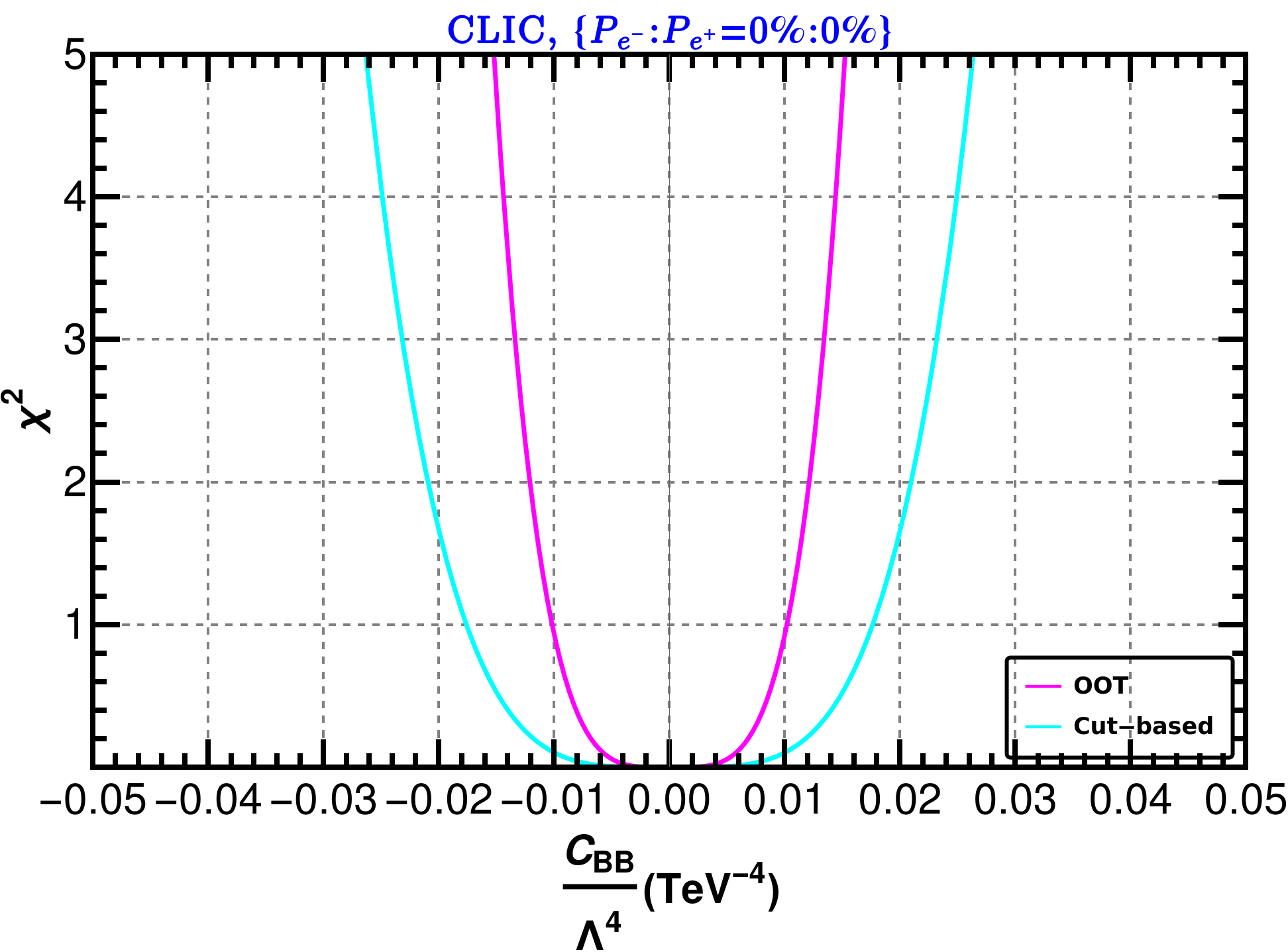}
	\includegraphics[height=4.5cm, width=5.1cm]{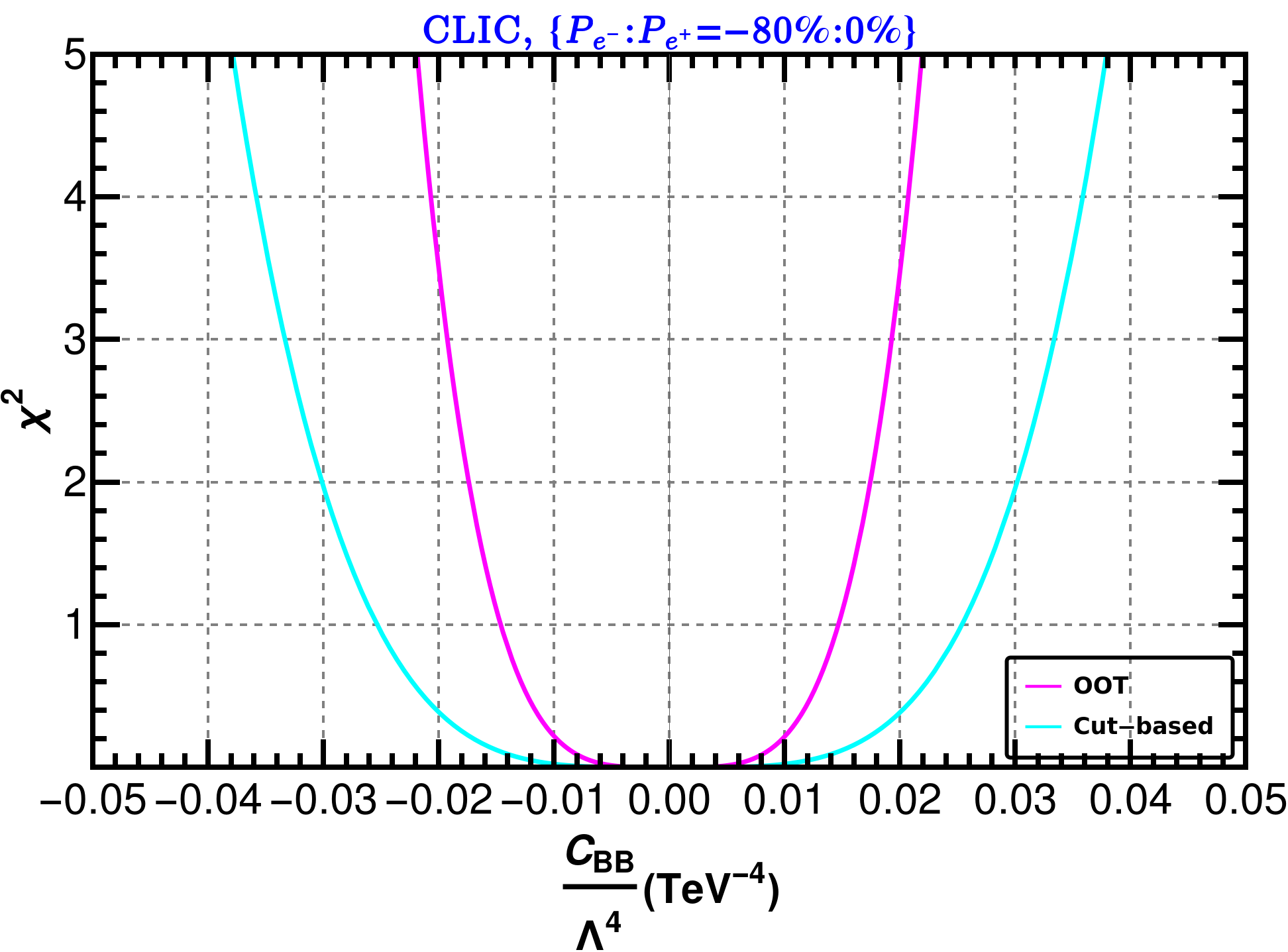}
	\includegraphics[height=4.5cm, width=5.1cm]{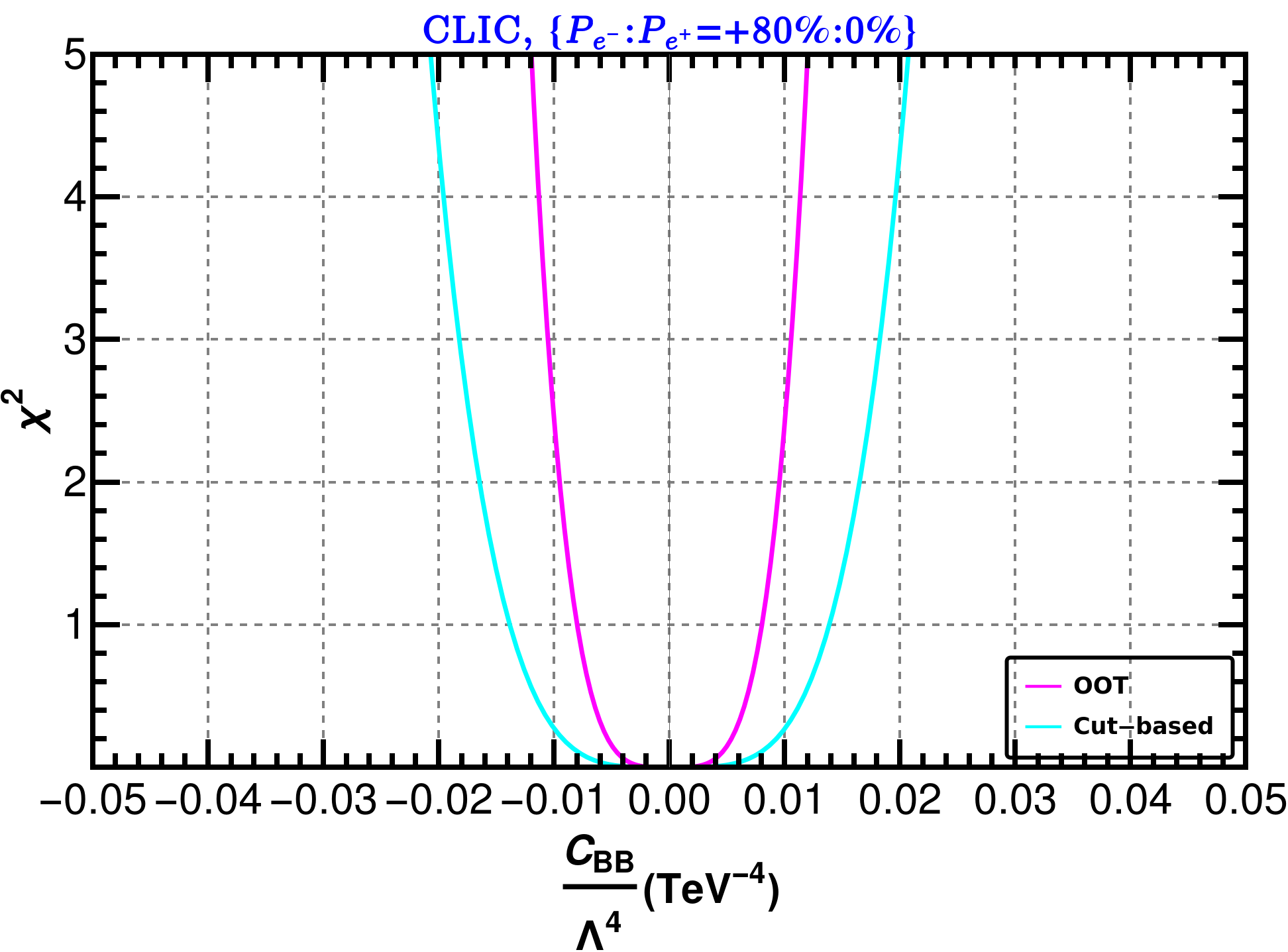}
	$$
	\caption{Comparison of the sensitivity of $\left(\frac{C_{BB}}{\Lambda^4}\right)$ for OOT and cut-based analysis. Beam polarization information is written in the captions of the figures. Top panel: ILC; Bottom panel: CLIC.} 	
	\label{fig:ootvscut}
\end{figure}

In this section, we investigate the estimation of the sensitivity of aNTGCs through cut-based analysis and compare the results derived from OOT. The $\chi^2$ function for cut-based analysis is given by,
\begin{equation}
	\chi^2 =\sum^{\rm{bins}}_{j} \left(\frac{N_j^{\tt obs}-N_j^{\tt theo}(g_i)}{\Delta N_j}\right)^2,
\end{equation}
where $N_j^{\tt obs}$ and $N_j^{\tt theo}$ are the number of events from observation and theory in the $\rm{j^{\tt th}}$ bins of differential cross-section distribution after applying all the cuts described in section~\ref{sec:col}. The statistical uncertainty in $\rm{j^{\tt th}}$ bin $\left(\Delta N_j \right)$ is $ = \sqrt{N^{\tt obs}_j}$, assuming the number of events in each bin follows Poisson distribution. Now using cut-based analysis, we briefly discuss the measurement of only one dim-8 effective coupling. Therefore,  we consider the most constrained dim-8 effective coupling $i.e.$ $\left(\frac{C_{BB}}{\Lambda^4}\right)$ among these four couplings. We again take up the design details of two linear colliders listed in Table~\ref{tab:design} and show the 95\% C.L. statistical limit of $\left(\frac{C_{BB}}{\Lambda^4}\right)$ by the cyan color in the figure~\ref{fig:ootvscut}. The 95\% C.L limits are tabulated in Table~\ref{tab:95clcut}. The estimation of the sensitivity via OOT of $\left(\frac{C_{BB}}{\Lambda^4}\right)$  is shown in magenta in the figure~\ref{fig:ootvscut}. Now, if we compare the sensitivity of dim-8 NP coupling, OOT outperforms cut-based analysis by a factor of 1.8. A detailed analysis in this regard is our future case of study. 

\begin{table}[htb!]
	\centering
	\begin{tabular}{| c | c |  c | c |  c | c | c | c   c | c  c | c  c| c  c| c c| } 
		\hline
		\multicolumn{1}{|c}{} &
		\multicolumn{6}{|c|}{$95\%$ C.L. limit} \\
		\cline{2-7}
		\multicolumn{1}{|c}{Coupling} &
		\multicolumn{3}{|c}{ILC} & 
		\multicolumn{3}{|c|}{CLIC} \\
		\cline{2-7}
		\multicolumn{1}{|c|}{$\rm (TeV^{-4})$}&
		\multicolumn{1}{c|}{$P_{e^\pm} = ^{00\%}_{00\%}$} &
		\multicolumn{1}{c}{$P_{e^\pm} = ^{ +30\%}_{-80\%}$}&
		\multicolumn{1}{|c|}{$P_{e^\pm} = ^{-30\%}_{+80\%}$}&
		\multicolumn{1}{c}{$P_{e^\pm} = ^{00\%}_{00\%}$}&
		\multicolumn{1}{|c}{$P_{e^\pm} = ^{+00\%}_{-80\%}$}&
		\multicolumn{1}{|c|}{$P_{e^\pm} = ^{+00\%}_{+80\%}$}\\
		\hline
		\multirow{2}*{$\frac{C_{BB}}{\Lambda^4}$}&  $+0.401$ & $+0.540$ &  $+0.285$ &  $+0.025$ &  $+0.036$  & $+0.020$ \\
		& $-0.401$ & $-0.540$ & $-0.285$ & $-0.025$ & $-0.036$ & $-0.020$ \\
		\hline
	\end{tabular}
	\caption{95\% C.L. limit of $\left(\frac{C_{BB}}{\Lambda^4}\right)$ through cut-based analysis at ILC and CLIC for different beam polarization combinations.}
	\label{tab:95clcut}
\end{table}

We would like to reiterate that our analysis only involves non-zero effects from dimension-8 operators. However, the inclusion of loop-induced NP effects from dimension-6 operators contribute to $Z \gamma$ production cross-section with ${\lsim 50\%}$ of that of the tree-level results at CLIC ($\sqrt{s}=3$ TeV and $\Lambda = 3.2$ TeV) and ${\lsim 8\%}$ of the tree-level result at ILC ($\sqrt{s} = 1$TeV and $\Lambda$=1.3 TeV). Although the dominant contribution comes from dimension-8 operators, for especially the CLIC analysis, the effect of loop-effects from dimension-6 operators cannot be neglected. Having said that, in this work, we have obtained the maximum sensitivity that can be achieved for dimension-8 couplings (individual or pair-wise) while all the other couplings are assumed to be zero. The same approach has been taken in \cite{ATLAS:2018nci}. The inclusion of all dimension-6 and 8 terms in a marginalized analysis will definitely weaken the limits obtained thus far. In fact, obtaining the best possible sensitivity for all the dimension-6 and -8 operators and taking in consideration all the relevant experimental results, can be the topic of a more elaborate future study.

\section{Conclusion}
\label{sec:con}

In this work, we have considered dimension-8 operators, that are consistent with SM gauge group, using SMEFT framework, which can give rise to anomalous neutral triple gauge couplings, namely $ZZ\gamma$ and $Z\gamma\gamma$. For this purpose, we have considered $Z \gamma$ production at the linear $e^+e^-$ colliders, the most suitable environment to study precision observables to probe physics beyond SM. We have analyzed mono-photon + missing energy as the final state signal and used useful kinematical cuts to segregate the signal from non-interfering SM backgrounds. In this context, missing energy ($\slashed{E}$) and missing transverse energy ($\slashed{E_T}$) play important role.

Next we have adopted optimal observable technique to find optimal sensitivity that can be achieved in the measurement of aNTGCs at the experiments. We have considered differential cross-section of $Z\gamma$ production as the observable of optimal observable technique to estimate statistical limit on the aNTGCs which ensures that statistical limit are optimal in the sense that the covariance matrix is minimized. Using OOT, we have estimated the 95\% C.L. limit on the dimension-8 aNTGCs. On one hand, we have performed individual one-dimensional analysis for all the couplings, on the other hand, the two-parameter analysis has also been performed in order to explore non-trivial correlation between different NP couplings. A comparative analysis of NP couplings between our predictions for future colliders, namely, ILC and CLIC and the most stringent limits obtained so far from ATLAS has been performed. While both linear colliders outperforms ATLAS with certain polarization combinations, CLIC 
offers at least ten times improvement over ILC sensitivities. For all the cases, the non-trivial effect of beam polarization can be noted. It is worthwhile to mention that different combination of polarized beams are sensitive to different couplings. Considering the choice of beam polarization that provides the most stringent limit at CLIC, we have observed that the optimal limit on dimension-8 aNTGCs are at least fifteen times better than the current experimental limit.

\acknowledgments

The authors would like to thank Subhaditya Bhattacharya and Jose Wudka for useful discussions.

\bibliographystyle{JHEP}
\bibliography{ref-ntgc}
\end{document}